\documentclass[pre,reprint]{revtex4-1}

\usepackage[utf8x]{inputenc}

\usepackage{mathtools,enumerate}
\usepackage[mathscr]{eucal}
\usepackage{amsmath,amssymb}
\usepackage[pdftex]{graphicx}
\usepackage{xcolor}

\def\ess{\mathscr{\varsigma}}

\begin{document}
\title{Approximate probability distributions of the master equation}
\author{Philipp Thomas}
\email{philipp.thomas@ed.ac.uk}
\affiliation{School of Mathematics and School of Biological Sciences, University of Edinburgh}
\author{Ramon Grima}
\email{ramon.grima@ed.ac.uk}
\affiliation{School of Biological Sciences, University of Edinburgh}

\begin{abstract}
Master equations are common descriptions of mesoscopic systems. Analytical solutions to these equations can rarely be obtained. We here derive an analytical approximation of the time-dependent probability distribution of the master equation using orthogonal polynomials. The solution is given in two alternative formulations: a series with continuous and a series with discrete support both of which can be systematically truncated. While both approximations satisfy the system size expansion of the master equation, the continuous distribution approximations become increasingly negative and tend to oscillations with increasing truncation order. In contrast, the discrete approximations rapidly converge to the underlying non-Gaussian distributions. The theory is shown to lead to particularly simple analytical expressions for the probability distributions of molecule numbers in metabolic reactions and gene expression systems.
\end{abstract}

\maketitle

\section{Introduction}

Master equations are commonly used to describe fluctuations of particulate systems. In most instances, however, the number of reachable states is so large that their combinatorial complexity prevents one from obtaining analytical solutions to these equations. Explicit solutions are known only for certain classes of linear birth-death processes \cite{jahnke2007}, under detailed balance conditions \cite{haken1974,*vanKampen1976}, or for particularly simple examples in stationary conditions \cite{mazo1975,*peccoud1995,*bokes2012}. Considerable effort has been undertaken to approximate the solution of the master equation under more general conditions including time-dependence and conditions lacking detailed balance \cite{gortz1976,*haag1979,schnoerr2014}. 

A common technique addressing this issue was given by van Kampen in terms of the system size expansion (SSE) \cite{van1976,*vanKampen}. The method assumes the existence of a specific parameter, termed the system size, for which the master equation approaches a deterministic limit as its value is taken to infinity. The leading order term of this expansion describes small fluctuations about this limit in terms of a Gaussian probability density, called the linear noise approximation (LNA). This approximation has been widely applied in biochemical kinetics \cite{ElfEhrenberg}, but also in the theory of polymer assembly \cite{melbinger2012,*szavits2014}, epidemics \cite{rozhnova2009}, economics \cite{aoki2001}, and machine learning \cite{heskes1994}. The benefit of the LNA is that it yields generic expressions for the probability density. Its deficiency lies in the fact that, strictly speaking, it is valid only in the limit of infinite system size. Hence one generally suspects that its predictions become inaccurate when one studies fluctuations that are not too small compared to the mean and therefore implying non-Gaussian statistics.

Higher order terms in the SSE have been employed to calculate non-Gaussian corrections to the LNA for the first few moments \cite{grima2009,*thomas2010,grima2011,cianci2012,*thomas2014}; alternative methods are based on moment closure \cite{engblom2006,*grima2012,*ale2013}. It is however the case that the knowledge of a limited number of moments does not allow to uniquely determine the underlying distribution functions. Reconstruction of the probability distribution therefore requires additional approximations such as the maximum entropy principle \cite{sotiropoulos2011,*smadbeck2013,*alexander2014} or the truncated moment generating function \cite{cianci2011} which generally yield different results. While the accuracy of these repeated approximations remains unknown, analytical expressions for the probability density can rarely be obtained, or might not even exist \cite{CoverThomas}. {A systematic investigation of the distributions implied by the higher order terms in the SSE, without resorting to moments, is therefore still missing.}

{
We here analytically derive, for the first time, a closed-form series expansion of the probability distribution underlying the master equation. We proceed by outlining the expansion of the master equation in Section I and briefly review the solution of the leading order terms given by the LNA in Section II. While commonly the SSE is truncated at this point, we show that the higher order terms can be obtained using an asymptotic expansion of the continuous probability density. The resulting series is given in terms orthogonal polynomials and can be truncated systematically to any desired order in the inverse system size. Analytical expressions are given for the expansion coefficients.}  

{
Thereby we establish two alternative formulations of this expansion: a continuous and a discrete one both satisfying the expansion of the master equation. We show that for linear birth-death processes, the continuous approximation often fails to converge reasonably fast. In contrast, the discrete approximation introduced in Section III accurately converges to the true distribution with increasing truncation order. In Section IV, we show that for nonlinear birth-death processes, renormalization is required for achieving rapid convergence of the series. Our analysis is motivated by the use of simple examples throughout. Using a common model of gene expression, we conclude in Section VI that the new method allows to predict the full time-dependence of the molecule number distribution.}

\section{System size expansion}

As a starting point, we focus on the master equation formulation of biochemical kinetics. We therefore consider a set of $R$ chemical reactions involving a single species confined in a well-mixed volume $\Omega$. Note that for chemical systems the system size coincides with the reaction volume. We denote by $S_{r}$ the net-change in the molecule numbers in the $r^{th}$ reaction and by $\gamma_r(n,\Omega)$ the probability per unit time for this reaction to occur. The probability of finding $n$ molecules in the volume $\Omega$ at time $t$, denoted by $P({n},t)$, then obeys the master equation

\begin{equation}
 \label{eqn:CME}
 \frac{\mathrm{d} P({n},t)}{\mathrm{d} t} = \sum_{r=1}^{R} \biggl(
 E^{-S_{r}} - 1 \biggr)
\gamma_r\left(n,\Omega\right) P({n},t),
 \end{equation}
where $E^{-S_{r}}$ is the step operator defined as $E^{-S_{r}}g(n)=g(n-S_{r})$ for any function $g(n)$ of the molecule numbers \cite{vanKampen}. Note that throughout the article, deterministic initial conditions are assumed.
The system size expansion now proceeds by separating the instantaneous concentration into a deterministic part, given by the solution of the rate equations $[X]$, and a fluctuating part $\epsilon$, 

\begin{align}
 \label{eqn:ansatz}
 \frac{n}{\Omega}=[X] + \Omega^{-1/2}\epsilon, 
\end{align}
which is van Kampen's ansatz. The expansion of the master equation can be summarized in three steps:
\paragraph*{(i)} Using Eq. (\ref{eqn:ansatz}) one expands the step operator
\begin{align}
 \label{eqn:sse_stepi}
 &E^{-S_r}\gamma_r\left(n,\Omega\right) P({n},t)  
  =\gamma_r\left(n-S_r,\Omega\right) P(n-S_r,t) \notag \\
  &= e^{-\Omega^{-1/2}S_r\partial_\epsilon}\gamma_r(\Omega[X]+\Omega^{1/2}\epsilon,\Omega) P(\Omega[X]+\Omega^{1/2}\epsilon,t),
\end{align}
where $\partial_\epsilon$ denotes $\frac{\partial}{\partial\epsilon}$.
\paragraph*{(ii)} {Next, the probability for the molecule numbers is cast into a probability density $\Pi(\epsilon,t)$ for the fluctuations using van Kampen's ansatz,
\begin{align}
 \label{eqn:sse_stepia}
 \Pi(\epsilon,t) = \Omega^{1/2} P(\Omega[X]+\Omega^{1/2}{\epsilon},t),
\end{align}
which is essentially a change of variables. Note that this step implicitly assumes a continuous approximation $\Pi(\epsilon,t)$ of the probability distribution as thought in the original derivation of van Kampen \cite{vanKampen}.}
\paragraph*{(iii)} It remains to expand the propensity about the deterministic limit
\begin{align}
 \gamma_r(\Omega[X]+\Omega^{1/2}\epsilon,\Omega)
 = \sum_{k=0}^\infty \Omega^{-k/2} \frac{\epsilon^k}{k!} \frac{\partial^k \gamma_r(\Omega[X],\Omega)}{\partial [X]^k}. 
\end{align}
Note that $\gamma_r(\Omega[X],\Omega)$ is just the propensity evaluated at the macroscopic concentration and hence it must depend explicitly on $\Omega$. We assume that the propensity possesses a power series in the inverse volume
\begin{align}
 \label{eqn:sse_stepiii}
 \gamma_r(\Omega[X],\Omega)=\Omega\sum_{s=0}^\infty \Omega^{-s} f^{(s)}_r\left([X]\right).
\end{align}
{For mass-action kinetics, for instance, the propensity is given by 
$\gamma_r(n,\Omega) =  \Omega^{1-\ell_r} {k_r} \ell_r! \binom{n}{\ell_r}$, 
where $\ell_r$ is the reaction order of the $r^{th}$ reaction. 
Using the Taylor expansion of the binomial coefficient, we have $f^{(0)}_r([X])=k_r [X]^{\ell_r}$, $f^{(s)}_r([X])=k_r[X]^{\ell_r-s} \mathcal{S}_{\ell_r,\ell_r-s}$, and $f^{(s)}_r=0$ for $s\geq\ell_r$, where $\mathcal{S}$ denotes the Stirling numbers of the first kind. Note also that effective propensities being deduced from mass action kinetics have an expansion similar to Eq. (\ref{eqn:sse_stepiii}). The Michaelis-Menten propensity $\gamma_r(n,\Omega) =  \Omega k_r \frac{n}{n+K\Omega}$ \cite{thomas2011}, for instance, has $f^{(0)}_r([X])=k_r \frac{[X]}{[X]+K}$ and $f^{(s)}_r([X])=0$ for $s>0$.}


Substituting now Eqs. (\ref{eqn:sse_stepi}-\ref{eqn:sse_stepiii}) into Eq. (\ref{eqn:CME}) and rearranging the result in powers of $\Omega^{-1/2}$, we find

\begin{align}
 \label{eqn:CMEexp}
 \biggl(&\frac{\partial}{\partial t}  -  \Omega^{1/2}\frac{\mathrm{d} [X]}{\mathrm{d} t} \frac{\partial}{\partial \epsilon}  \biggr)\Pi(\epsilon,t) \notag\\ 
 =& \biggl(- \Omega^{1/2}\sum_{r=1}^R S_r f_r^{(0)}([X]) \frac{\partial}{\partial \epsilon}+\sum_{k=0}^N\Omega^{-k/2}\mathcal{L}_k \biggr) \Pi(\epsilon,t) \notag\\&+ O(\Omega^{-(N+1)/2}).
\end{align}
Equating terms to order $\Omega^{1/2}$ yields the deterministic rate equation 

\begin{align}
 \label{eqn:REs}
 \frac{\mathrm{d} [X]}{\mathrm{d} t}=\sum_{r=1}^R S_r f_r^{(0)}([X]).
\end{align}
{
The higher order terms in the expansion of the master equation can be written out explicitly
\begin{align}
 \mathcal{L}_k = \sum_{s=0}^{\lceil k/2\rceil} \sum_{p=1}^{k-2(s-1)} \frac{\mathcal{D}_{p,s}^{k-p-2(s-1)}}{p!(k-p-2(s-1))!} (-\partial_\epsilon)^p \epsilon^{k-p-2(s-1)},
\end{align}
where $\lceil \cdot \rceil$ denotes the ceiling value and the coefficients are given by
\begin{align}
 \label{eqn:SSEcoeffs}
 \mathcal{D}_{p,s}^{q}= \sum_{r=1}^R (S_r)^p \partial_{[X]}^q f_r^{(s)}([X]),
\end{align}
which depend explicitly on the solution of the rate equation (\ref{eqn:REs}). Note that in the following the abbreviation $\mathcal{D}_p^{q}=\mathcal{D}_{p,0}^{q}$ is used.}

\section{Expansion of the continuous probability density}

We here study the time-dependent solution of the partial differential equation approximation of the master equation, Eq. (\ref{eqn:CMEexp}). We therefore expand the probability density of Eq. (\ref{eqn:sse_stepia}),

\begin{align}
 \label{eqn:asymptoticExpansion}
 \Pi(\epsilon,t)=\sum_{j=0}^N \Omega^{-j/2} \pi_j(\epsilon,t) + O(\Omega^{-(N+1)/2}),
\end{align}
which also allows the expansion of the time-dependent moments to be deduced in closed-form. Finally we recover the stationary solution as a particular case.

\subsection{Linear Noise Approximation}

Substituting Eq. (\ref{eqn:asymptoticExpansion}) into Eq. (\ref{eqn:CMEexp}) and equating terms to order $O(\Omega^{0})$ we find 

\begin{align}
\left(\frac{\partial}{\partial t} - \mathcal{L}_0  \right) \pi_{0} = 0,
\end{align}
where $\mathcal{L}_0=-\partial_\epsilon \mathcal{J} \epsilon + \frac{1}{2} \partial_\epsilon^2 \mathcal{D}_2^0$ is a Fokker-Planck operator with linear coefficients, and $\mathcal{J}=\mathcal{D}_1^1$ is the Jacobian of the rate equation.
The probability density of fluctuations about the macroscopic concentration, described by $\epsilon$, is given by a centered Gaussian 

\begin{align}
 \label{eqn:pi0}
 \pi_0(\epsilon,t)=\frac{1}{\sqrt{2\pi\sigma^2(t)}} \exp\left({-\frac{\epsilon^2}{2\sigma^2(t)}}\right), 
\end{align}
which acquires time-dependence via its variance $\sigma^2(t)$. The latter satisfies

\begin{align}
 \label{eqn:lyapunov}
 \frac{\partial \sigma^2}{\partial t}  = 2 \mathcal{J}(t) \sigma^2 + \mathcal{D}_2^0(t),
\end{align}
which is the familiar LNA result \cite{vanKampen}. In the following we will drop the time-dependence of the coefficients for convenience of notation.

\subsection{Higher order terms}

Substituting Eq. (\ref{eqn:asymptoticExpansion}) into Eq. (\ref{eqn:CMEexp}), rearranging the remaining terms, and equating terms to order $\Omega^{-j/2}$, we find

\begin{align}
  \label{eqn:systemofequations}
  \biggl(\frac{\partial}{\partial t}-\mathcal{L}_0\biggr) \pi_{j}(\epsilon,t) \notag
  &= \mathcal{L}_1 \pi_{j-1} + \ldots + \mathcal{L}_j \pi_0 \\
  &= \sum_{k=1}^j \mathcal{L}_k \pi_{j-k}(\epsilon,t).
\end{align}
This system of partial differential equations can be solved using the eigenfunction approach. We consider 

\begin{align}
 \label{eqn:eigenvalueproblem}
 \left(\frac{\partial}{\partial t}-\mathcal{L}_0\right) \Psi_m  = \lambda_m  \Psi_m,
\end{align}
which is solved by $\lambda_m=-m\mathcal{J}$ and $\Psi_m=\psi_m(\epsilon,t)\pi_0(\epsilon,t)$ with

\begin{align}
 \label{eqn:eigenfunctions}
 \psi_m(\epsilon,t)=\pi_0^{-1}(-\partial_\epsilon)^m \pi_0 = \frac{1}{\sigma^m} H_m\left(\frac{\epsilon}{\sigma}\right).
\end{align}
The functions $H_m$ denote the Hermite orthogonal polynomials which are given explicitly in Appendix \ref{app:Hermite}. To verify the solution of the eigenvalue problem, we set $\Psi_{m+1}=(-\partial_\epsilon)\Psi_{m}$ and observe that $(\partial_t-\mathcal{L}_0)\Psi_{m+1}=-\mathcal{J}\Psi_{m+1}-\partial_\epsilon(\partial_t-\mathcal{L}_0)\Psi_{m}$. Using this in Eq. (\ref{eqn:eigenvalueproblem}), we obtain $\lambda_{m+1}=(-\mathcal{J}+\lambda_{m})$ from which the result follows because $\lambda_0=0$ and $\Psi_0=\pi_0$.

Using the completeness of the eigenfunctions, we can write $\pi_j(\epsilon,t)=\sum_{m=0}^{\infty} a_m^{(j)}(t) \psi_m({\epsilon},t)\pi_0(\epsilon,t)$. We verify in Appendix \ref{app:integral} that the $j^{th}$ order term in the expansion involves only the first $N_j=3j$ eigenfunctions. The continuous SSE approximation is consequently given by the asymptotic expansion

\begin{align}
 \label{eqn:HermiteExpansion}
 \Pi(\epsilon,t) =& \pi_0(\epsilon,t)\biggl( 1+\sum_{j=1}^{N} \Omega^{-j/2}\sum_{m=1}^{N_j} a_m^{(j)}(t) \psi_m\left({\epsilon},t\right) \biggr) 
 \nonumber\\&+ O(\Omega^{-(N+1)/2}),
\end{align}
for which the coefficients can be determined using the orthogonality of the functions $\psi_m$, i.e., $\frac{\sigma^{2n}}{n!}\int \mathrm{d} \epsilon \, \psi_n(\epsilon,t) \psi_m(\epsilon,t) \pi_0(\epsilon,t)=\delta_{m,n}$

\subsection{The equation for the expansion coefficients}

The coefficients $a_n^{(j)}$ are now determined by 
inserting the expansion of $\pi_j$ into Eq. (\ref{eqn:systemofequations}), multiplying the result by $\frac{\sigma^{2n}}{n!}\int \mathrm{d} \epsilon \, \psi_n(\epsilon,t)$, and performing the integration. Using Eq. (\ref{eqn:eigenvalueproblem}), the left hand side of Eq. (\ref{eqn:systemofequations}) becomes

\begin{align}
 \label{eqn:lhs}
 \frac{\sigma^{2n}}{n!}\sum_m & \int \mathrm{d} \epsilon \, \psi_n\left({\epsilon},t\right) \left( \frac{\partial}{\partial t}
-\mathcal{L}_0\right) a_m^{(j)} \psi_m\left({\epsilon},t\right) \pi_0(\epsilon,t) \notag \\
 =& \left( \frac{\partial}{\partial t}- n \mathcal{J}\right) a_n^{(j)}.
\end{align}
The calculation of terms in the summation on the right hand side of Eq. (\ref{eqn:systemofequations}) is greatly simplified by defining the integral 

\begin{equation}
\label{eqn:integraldef}
\mathcal{I}^{\alpha\beta}_{mn} = \frac{\sigma^{2n}}{n!\alpha!\beta!} \int \mathrm{d}\epsilon \, \psi_n({\epsilon},t) (-\partial_\epsilon)^\alpha \epsilon^\beta \psi_m(\epsilon,t) \pi_0(\epsilon,t),
\end{equation}
which yields

\begin{align}
 \label{eqn:rhs}
 \frac{\sigma^{2n}}{n!}&\int d \epsilon \psi_n(\epsilon,t) \mathcal{L}_k \psi_m(\epsilon,t)\pi_0(\epsilon,t) \notag\\
 =& \sum_{s=0}^{\lceil k/2\rceil} \sum_{p=1}^{k-2(s-1)} \mathcal{D}_{p,s}^{k-p-2(s-1)} \mathcal{I}_{mn}^{p,k-p-2(s-1)}.
\end{align} 
Using Eqs. (\ref{eqn:lhs}) and (\ref{eqn:rhs}) in Eq. (\ref{eqn:systemofequations}), we find that the coefficients satisfy the following set of ordinary differential equations

\begin{align}
 \label{eqn:coefficientEquations}
 &\left( \frac{\partial}{\partial t}- n \mathcal{J} \right) a_n^{(j)} =\nonumber\\
 &\sum_{k=1}^j \sum_{m=0}^{N_{j-k}} a_m^{(j-k)} \sum_{s=0}^{\lceil k/2\rceil} \sum_{p=1}^{k-2(s-1)} \mathcal{D}_{p,s}^{k-p-2(s-1)} \mathcal{I}_{mn}^{p,k-p-2(s-1)},
\end{align}
where we have assumed $a_n^{(j)}=0$ for $n>N_j$. Explicitly, the non-zero integrals are given by

\begin{align}
 \label{eqn:integral}
 \mathcal{I}^{\alpha\beta}_{mn}
   =& \frac{\sigma^{\beta-\alpha+n-m}}{\alpha!}\sum_{s=0}^{\min({n-\alpha,m})}\binom ms \times\notag \\& \frac{(\beta+\alpha+2s-(m+n)-1)!!}{(\beta+\alpha+2s-(m+n))!(n-\alpha-s)!},
\end{align}
and zero for odd $(\alpha+\beta)-(m+n)$. {Here $(2k-1)!! = \frac{(2k)!}{2^k k!}$ is the double factorial. Along with Eq. (\ref{eqn:integral}), in Appendix \ref{app:integral} we verify the following two important properties of the asymptotic series solution given deterministic initial conditions:
(i) We have $N_j=3j$ and hence Eq. (\ref{eqn:coefficientEquations}) indeed yields a finite number of equations, and
(ii) $a_n^{(j)}$ vanishes for all times when $(n+j)$ is odd.}

{Finally, we note that $\mathcal{D}_{p,s}^q$ and $\mathcal{I}_{mn}^{pq}$ are generally time-dependent because they are functions of the solution of the rate equation and the LNA variance. Explicit expressions for the approximate probability density can now be evaluated to any desired order.}

\subsection{Moments of the distribution}

The solution for the probability density enables one to derive closed-form expressions for the moments. These are obtained by multiplying Eq. (\ref{eqn:HermiteExpansion}) by $\int \mathrm{d} \epsilon \,\epsilon^\beta$ and performing the integration using Eq. (\ref{app:HermiteMoments}) of Appendix \ref{app:integral}. We find

\begin{align}
 \label{eqn:MomentsFromCoeffs}
 \langle\epsilon^\beta\rangle
 = \sum_{j=0}^N \Omega^{-j/2} 
  \sum_{k=0}^{\lfloor\beta/2\rfloor} &\frac{\beta!}{2^k k!} \sigma^{2k} a_{\beta-2k}^{(j)} 
  + O(\Omega^{-(N+1)/2}),
\end{align}
where $a_0^{(j)}=\delta_{0,j}$ and $\lfloor \cdot \rfloor$ denotes the floor value. In particular, it follows that mean and variance are given by $\langle\epsilon\rangle=\sum_{j=1}^{N} \Omega^{-j/2} a_1^{(j)} + O(\Omega^{-(N+1)/2})$ and $\langle\epsilon^2\rangle=\sigma^2+2 \sum_{j=1}^{N} \Omega^{-j/2}	 a_2^{(j)} + O(\Omega^{-(N+1)/2})$. 

It is now evident that the coefficients of the expansion are intricately related to the system size expansion of the distribution moments. Naturally, one may seek to invert this relation. Indeed, as we show in Appendix \ref{app:problemofmoments}, given the expansion for a finite set moments, the coefficients in Eq. (\ref{eqn:HermiteExpansion}) can be uniquely determined. In particular, to construct the probability density to order $\Omega^{-j/2}$ one requires the expansion of the first $3j$ moments to the same order. Thus the problem of moments provides an equivalent route of systematically constructing solutions to the master equation.

\subsection{Solution in stationary conditions}

\label{sec:steadystate}

Of particular interest is the expansion of the probability density under stationary conditions. Implicitly, we assume here that the rate equation, Eq. (\ref{eqn:REs}), has a single asymptotically stable fixed point, and hence the LNA variance is given by $\sigma^2=\mathcal{D}_{2}^0/(-2\mathcal{J})$. Setting the time-derivative on the left hand side of Eq. (\ref{eqn:coefficientEquations}) to zero, we find that the coefficients of Eq. (\ref{eqn:HermiteExpansion}) can be expressed in terms of lower order ones

\begin{align}
 \label{eqn:coefficients}
 a_n^{(j)} = -&\frac{1}{ n \mathcal{J}} \sum_{k=1}^j \sum_{s=0}^{\lceil k/2\rceil} \sum_{p=1}^{k-2(s-1)}\times  \notag \\ &  \mathcal{D}_{p,s}^{k-p-2(s-1)}\sum_{m=0}^{3({j-k})} a_m^{(j-k)} \mathcal{I}_{mn}^{p,k-p-2(s-1)}. 
\end{align}
For example, truncating after terms of order $\Omega^{-1}$, we obtain

\begin{align}
\label{eqn:O-1sol1}
 &\Pi(\epsilon) = \pi_0 (\epsilon) + \Omega^{-1/2} \left( a_1^{(1)} \psi_1\left(\epsilon\right)
    + a_3^{(1)} \psi_3\left(\epsilon\right)
    \right) \pi_0 (\epsilon) \notag \\
    & + \Omega^{-1} 
    \left( a_2^{(2)} \psi_2\left(\epsilon\right)
        +  a_4^{(2)} \psi_4\left(\epsilon\right)
        +  a_6^{(2)} \psi_6\left(\epsilon\right)
    \right) \pi_0 (\epsilon) \notag\\& + O(\Omega^{-3/2}).
\end{align}
The non-zero coefficients to order $\Omega^{-1/2}$ are given by
\begin{align}
\label{eqn:O-1sol2}
 a_1^{(1)} &= -\frac{\sigma ^2 \mathcal{D}_1^{2}}{2 \mathcal{J} }- \frac{ \mathcal{D}_{1,1}^{0}}{\mathcal{J} }, \notag \\
 a_3^{(1)} &= -\frac{\sigma^4 \mathcal{D}_1^{2}}{6 \mathcal{J}}-\frac{\sigma^2 \mathcal{D}_2^{1}}{6 \mathcal{J}}-\frac{\mathcal{D}_3^{0}}{18 \mathcal{J}},
\end{align}
while those to order $\Omega^{-1}$ are 
\begin{align}
\label{eqn:O-1sol3}
 a_2^{(2)}=&-a_1^{\text{(1)}}\left(
\frac{\mathcal{D}_{1,1}^{\text{0}} }{2 \mathcal{J} }
+\frac{ \mathcal{D}_2^{\text{1}}}{4 \mathcal{J} }
+\frac{3 \sigma^2 \mathcal{D}_1^{\text{2}}}{4 \mathcal{J}}
\right)
-a_3^{\text{(1)}}\frac{3  \mathcal{D}_1^{\text{2}}}{2 \mathcal{J}}
\notag\\&
-\frac{\mathcal{D}_{2,1}^{\text{0}}}{4 \mathcal{J}}-\frac{\sigma^2 \mathcal{D}_{1,1}^{\text{1}}}{2 \mathcal{J}}-\frac{\sigma^2  \mathcal{D}_2^{\text{2}}}{8 \mathcal{J}}-\frac{\sigma^4 \mathcal{D}_1^{\text{3}}}{4 \mathcal{J}}, \notag\\
 a_4^{(2)}=&
-a_1^{\text{(1)}}
\left(
\frac{\mathcal{D}_3^{\text{0}}}{24 \mathcal{J}}
+\frac{\sigma^2 \mathcal{D}_2^{\text{1}}}{8 \mathcal{J} }
+\frac{\sigma^4 \mathcal{D}_1^{\text{2}}}{8 \mathcal{J}}
\right)
-\frac{\mathcal{D}_4^{\text{0}}}{96 \mathcal{J}}
-\frac{ \sigma^2 \mathcal{D}_3^{\text{1}}}{24 \mathcal{J}}
\notag\\
 &
-\frac{ \sigma^4 \mathcal{D}_2^{\text{2}}}{16 \mathcal{J}}
-\frac{\sigma^6 \mathcal{D}_1^{\text{3}}}{24 \mathcal{J}}
-a_3^{\text{(1)}}\left(
 \frac{\mathcal{D}_{1,1}^{\text{0}}}{4 \mathcal{J} }
+\frac{3 \mathcal{D}_2^{\text{1}}}{8 \mathcal{J} }
+\frac{7 \sigma^2 \mathcal{D}_1^{\text{2}}}{8 \mathcal{J}}
\right),
\notag\\
 a_6^{(2)} =& \frac{1}{2} (a_3^{\text{(1)}})^2.
\end{align}
The accuracy of this distribution approximation is studied through an example in the following.
\begin{figure}
  \includegraphics[width=0.4\textwidth]{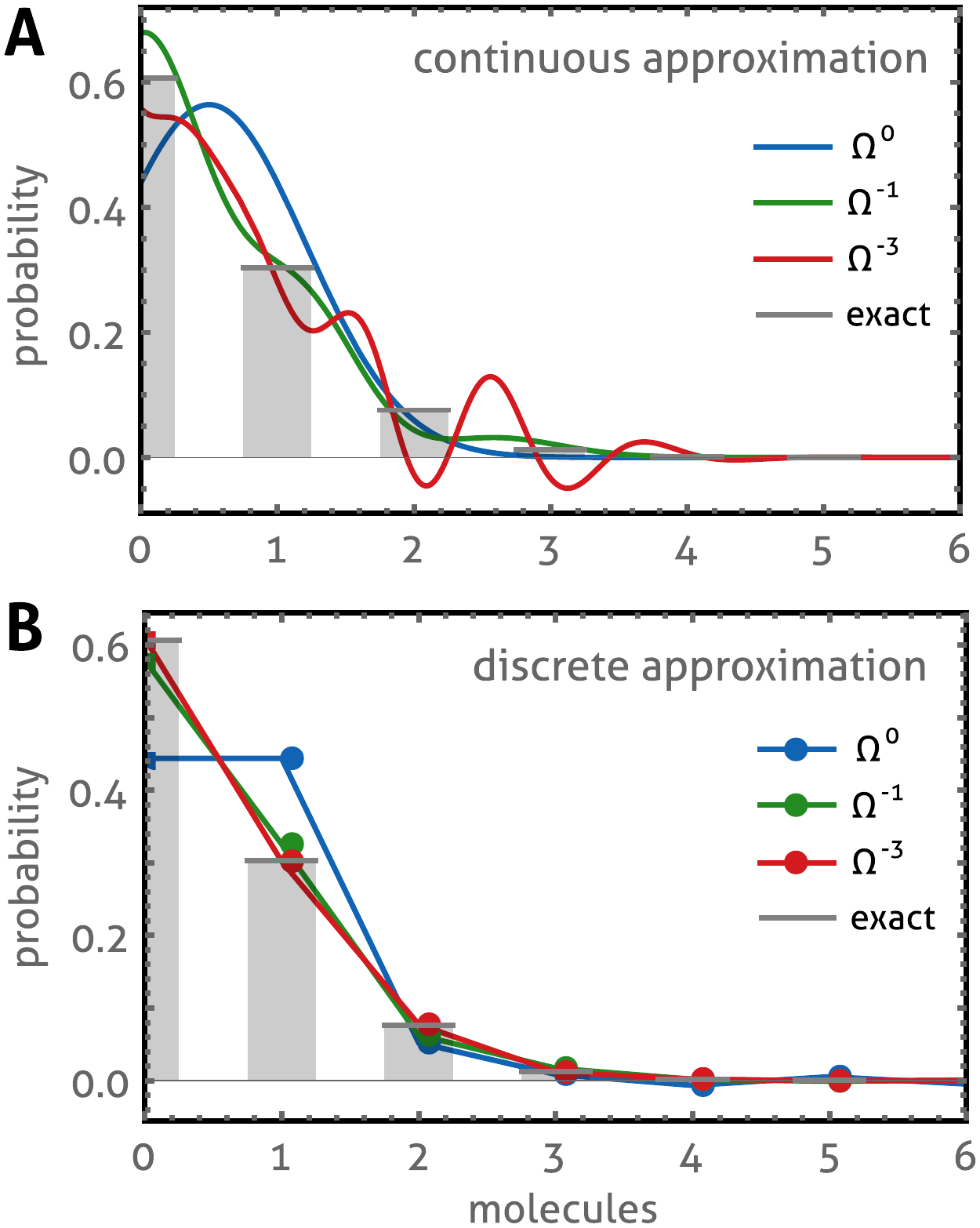} 
  \caption{(Color
 online) \textbf{Linear birth-death process.} We consider the reaction system (\ref{reaction:simple}) in stationary conditions. (A) We compare the exact Poisson distribution (gray) to the continuous SSE approximation [Eq. (\ref{eqn:HermiteExpansion}) together with Eqs. (\ref{eqn:coefficients}) and (\ref{poisson:SSEcoeffs})] truncated after $\Omega^0$ (LNA, blue line), $\Omega^{-1}$ (green), and $\Omega^{-3}$-terms (red) for parameter values $k_0=0.5$, $k_1=1$ and $\Omega=1$ giving half a molecule on average. We observe that the continuous approximation becomes increasingly negative and tends to oscillations with increasing truncation order. (B) In contrast the discrete approximation shows no oscillations, and the overall agreement with the exact Poisson distribution (gray bars) improves with increasing truncation order. 
  }
  \label{fig:poisson}
\end{figure}

\subsection{The continuous approximation fails under low molecule number conditions}

We now study the SSE solution for a linear birth-death process, i.e., its propensities depend at most linearly on the molecular populations. Specifically, we consider the synthesis and decay of a molecular species $X$,

\begin{align}
 \label{reaction:simple}
 \varnothing \xrightleftharpoons[k_1]{k_0} X.
\end{align}
The master equation is constructed using $S_1=+1$, $\gamma_1=\Omega k_0$, $S_2=-1$, $\gamma_2=k_1 n$, and $R=2$ in Eq. (\ref{eqn:CME}). The exact stationary solution of the master equation is a Poisson distribution with mean $\Omega [X]$ where $[X]=k_0/k_1$. The coefficients in Eq. (\ref{eqn:SSEcoeffs}) are then given by 
\begin{align}
 \label{poisson:SSEcoeffs}
 \mathcal{D}_{n}^m= \delta_{m,0} k_0  + (-1)^n k_1 \left( \delta_{m,0} [X]+ \delta_{m,1}\right),
\end{align}
and $\mathcal{D}_{n,s}^{m}=0$ for $s>0$. The leading order corrections to the LNA given by Eqs. (\ref{eqn:O-1sol1}-\ref{eqn:O-1sol3}) lead to very compact expressions for the expansion coefficients and are given by
\begin{align}
 \label{poisson:compactExp}
 a_3^{\text{(1)}} = \frac{[X]}{6}, \ \
 a_4^{\text{(2)}} = \frac{[X]}{24}, \ \
 a_6^{\text{(2)}} = \frac{[X]^2}{72}
\end{align}
and $a_1^{\text{(1)}}= a_2^{\text{(2)}}=0$.

Though the continuous approximation is expected to perform well at large values of $\Omega$, we are particularly interested in its performance when the value of $\Omega$ is decreased. Since the expansion is carried out at constant average concentration, low values of $\Omega$ typically imply low numbers of molecules and non-Gaussian distributions. 
In Fig. \ref{fig:poisson}A we show that for parameters yielding half a molecule on average, the continuous approximation obtained in this section, given by Eq. (\ref{eqn:HermiteExpansion}) together with Eqs. (\ref{eqn:coefficients}) and (\ref{poisson:SSEcoeffs}), is unsatisfactory since as higher orders are taken into account, one observes large oscillations in the tails of the distribution. In the following section we show that the disagreement arises due to the assumption that the support of the distribution is continuous rather than discrete as implied by the master equation.

\section{Discrete approximation of the probability distribution}
\label{sec:discrete}

The aim of this paragraph is to establish a discrete formulation of the distribution approximations.. To clarify this issue, we note that the
exact characteristic function $G(k,t)=\sum_{n=0}^\infty e^{ikn} P(n,t)$ is a $2\pi$-periodic function, and hence can be inverted as follows

\begin{align}
 \label{eqn:CFinversion}
 P({n},t)=\int_{-\pi}^\pi\frac{\mathrm{d}k}{2\pi} e^{-ikn} G(k,t).
\end{align}
We now associate our continuous approximation, Eq. (\ref{eqn:HermiteExpansion}), with this characteristic function, i.e., $G(k,t)= \int_{-\infty}^\infty \mathrm{d}\epsilon\, e^{ik\Omega([X]+\Omega^{-1/2}\epsilon)} \Pi(\epsilon,t)$. Substituting this together with Eq. (\ref{eqn:asymptoticExpansion}) into Eq. (\ref{eqn:CFinversion}) one establishes a connection formula between these discrete and continuous approximations via the convolution

\begin{align}
 \label{eqn:connectionformula}
 P({n},t) =& \sum_{j=0}^{N} \Omega^{-j/2} \int_{-\infty}^\infty \mathrm{d}\epsilon \, K(n-\Omega[X]-\Omega^{1/2}\epsilon)  \pi_j(\epsilon,t) 
 \notag\\ &+ O(\Omega^{-(N+1)/2}),
\end{align}
with kernel $$K(s)=\int_{-\pi}^\pi \frac{\mathrm{d} k}{2\pi} e^{-iks} = \frac{\sin(\pi s)}{\pi s}.$$ The convolution can be used to define the derivatives of the discrete probability via

\begin{align}
 \label{eqn:derivativeformula}
 \partial_n P({n},t) = \int_{-\infty}^\infty \mathrm{d}\epsilon\, K(n-\Omega [X]-\Omega^{1/2}\epsilon) (\Omega^{-1/2}\partial_\epsilon) \Pi(\epsilon,t),
\end{align}
and hence it satisfies $E^{-S_j}P({n},t) = \int_{-\infty}^\infty \mathrm{d}\epsilon\, K(n-\Omega [X]-\Omega^{1/2}\epsilon) e^{-\Omega^{-1/2} \partial_\epsilon S_j}\Pi(\epsilon,t)$, as well as $\gamma_j(n,\Omega)P({n},t) = \int_{-\infty}^\infty \mathrm{d}\epsilon\, K(n-\Omega [X]-\Omega^{1/2}\epsilon) \gamma_j(\Omega[X]+\Omega^{1/2}\epsilon,\Omega) \Pi(\epsilon,t)$ for analytic $\gamma_j$. 
It then follows from the fact that $P(n,t)$ and $\Omega^{1/2}\Pi(\Omega^{-1/2}(n-\Omega[X]),t)$ have the same characteristic function expansion, that 
(i) both approximations possess the same asymptotic expansion of their moments, and that
(ii) they satisfy the same expansion of the master equation.

For example, to leading order $\Omega^0$, Eq. (\ref{eqn:connectionformula}) replaces the conventional continuous LNA estimate, $\pi_0$ given by Eq. (\ref{eqn:pi0}), with a discrete approximation
\begin{align}
 \label{eqn:P0}
 P_0(n,t)=\frac{1}{2}\frac{e^{-\frac{y^2}{2 \Sigma^2 }}}{\sqrt{2 \pi } {\Sigma}}{\left[\text{erf}\left(\frac{iy+\pi\Sigma^2 }{\sqrt{2}\Sigma}\right)-\text{erf}\left(\frac{iy- \pi \Sigma^2 }{\sqrt{2} {\Sigma}}\right)\right]},
\end{align}
where $y=n-\Omega[X]$, $\Sigma^2=\Omega\sigma^2$ is the LNA's estimate for the variance of molecule numbers, and $\text{erf}$ is the error function defined by $\operatorname{erf}(x) = \frac{2}{\sqrt\pi}\int_0^x e^{-t^2}\,\mathrm dt$. 

Associating the $\Omega^{-j/2}$-term of Eq. (\ref{eqn:HermiteExpansion}) with $\pi_j$ in Eq. (\ref{eqn:connectionformula}), higher order approximations can now be obtained from

\begin{align}
 \label{eqn:GeneralizedExpansion}
 P(n,t) &=  P_0(n,t) \notag\\
   &+ \sum_{j=1}^{N} \Omega^{-j/2}\sum_{m=1}^{3j} a_m^{(j)} \left(-\Omega^{1/2}\partial_n\right)^m P_0(n,t) 
\nonumber\\& + O(\Omega^{-(N+1)/2}).
\end{align}
The above follows from the definition of the eigenfunctions, Eq. (\ref{eqn:eigenfunctions}), and using the derivative property of the convolution given after Eq. (\ref{eqn:derivativeformula}). Note that the coefficients in this equation are exactly the same as given in Eq. (\ref{eqn:HermiteExpansion}) and hence are determined by Eq. (\ref{eqn:coefficientEquations}). One can verify two limiting cases: (i) as ${\Sigma \to 0}$ and $\Omega[X]$ being integer-valued, then $P_0(n)=K(n-\Omega[X])=\delta_{n,\Omega[X]}$ is just the Kronecker delta as required for deterministic initial conditions; (ii) as $\Omega\to\infty$ with $y/\Sigma$ constant, the probability distribution $P_0$ reduces to the density $\pi_0$ given by Eq. (\ref{eqn:pi0}) and hence it follows that in this limit the continuous and discrete series give the same results.

\subsection{The discrete approximation performs well for linear birth-death processes}

For the linear birth-death process in the previous section, in Fig. \ref{fig:poisson}B we show that the discrete approximation given by {Eq. (\ref{eqn:GeneralizedExpansion}) with Eq. (\ref{poisson:compactExp}) is in good agreement with the true distribution when truncated after terms of order $\Omega^{-1}$ and shows no oscillations.  This agreement is remarkable given the compact form of the solution given by Eq. (\ref{poisson:compactExp}) and (\ref{eqn:GeneralizedExpansion}). The approximation is almost indistinguishable from the exact result when the series is truncated after $\Omega^{-3}$-terms using Eqs. (\ref{eqn:coefficients}) and (\ref{poisson:SSEcoeffs}) in Eq. (\ref{eqn:GeneralizedExpansion}).} We hence conclude that the discrete series approximates better the underlying distribution of the master equation than the continuous approximation. 

\begin{figure*}
  \includegraphics[width=0.8\textwidth]{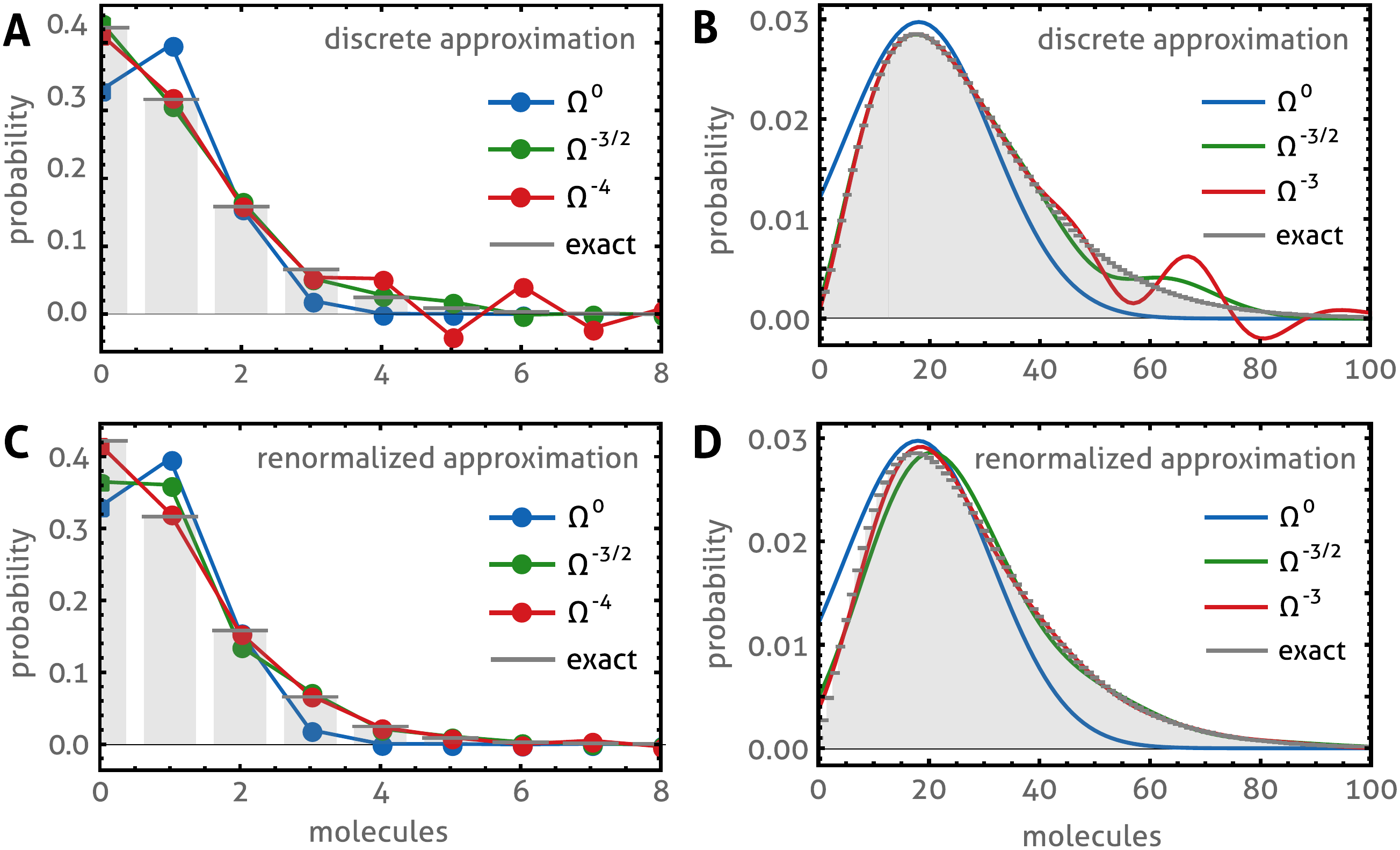} 
   \caption{(Color
 online) \textbf{Nonlinear birth-death process.} A metabolic reaction with Michaelis-Menten kinetics, scheme (\ref{reaction:MM}), is studied using the reduced model described in Sec. \ref{sec:MM}. The exact stationary distribution is a negative binomial (shown in gray). (A) The discrete SSE approximation given by Eq. (\ref{eqn:GeneralizedExpansion}) with Eq. (\ref{eqn:coefficients}) and (\ref{poisson:SSEcoeffs}) is shown in the low molecule number regime (${k_0}/{k_1}=0.25$, $1$ molecule on average) when truncated after $\Omega^0$ (blue), $\Omega^{-3/2}$ (green) and $\Omega^{-4}$-terms (red dots). We observe that the expansion tends to oscillations and negative values of probability as the truncation order is increased. (B) Similar oscillations are observed for moderate molecule numbers (${k_0}/{k_1}=0.9$, $27$ molecules on average) for the discrete series truncated after $\Omega^0$ (blue), $\Omega^{-3/2}$ (green) and $\Omega^{-3}$-terms (red lines).  In (C) and (D) we show the approximations corresponding to the same parameters used in (A) and (B), respectively, but obtained using the renormalization procedure given by Eq. (\ref{eqn:RenormalizedExpansion}) with Eq. (\ref{eqn:renormCoeffs}) as described in the main text. The renormalized approximations avoid oscillations and are in excellent agreement with the true probability distributions (gray bars). We note that for the cases (B) and (D) the continuous and discrete approximations give essentially the same result. The remaining parameters are given by $\Omega=10$ and $K=0.1$.}
  \label{fig:NB}
\end{figure*}

\subsection{The discrete approximation fails for non-linear birth processes}
\label{sec:MM}

Next, we turn our attention to the analysis of nonlinear birth-death processes, i.e., a process whose propensities depend nonlinearly on the number of molecules. A particular feature of such processes is that the LNA estimates for mean and variances are generally no longer exact, but agree with those of the true distribution only in the limit of large system size \cite{grima2011}. 

Exemplary, we here consider a simple metabolic reaction confined in a small subcellular compartment of volume $\Omega$ with substrate input, 

\begin{subequations}
\label{reaction:MM}
\begin{align}
 &\varnothing \xrightarrow[]{h_0} S, \label{MM:synthesis} \\
 &S + E \xrightleftharpoons[h_2]{h_1} C \xrightarrow[]{h_3} E. \label{MM:degr}
\end{align} 
\end{subequations}
The reactions describe the input of substrate molecules $S$ and their catalytic conversion by enzyme species $E$ via the enzyme-substrate complex $C$. The SSE of the average concentrations correcting the macroscopic rate equations have been extensively studied \cite{grima2009}.
Since our theory applies to a single species only, we here consider a reduced model in which reaction (\ref{MM:degr}) is modelled via an effective propensity: this gives $S_1=+1$, $\gamma_1=\Omega k_0$, and $S_2=-1$, $\gamma_2=\Omega k_1 \frac{n}{n+\Omega K}$. This simplification is valid when the enzyme-substrate association is in rapid equilibrium, which holds when $[E_T]\ll K$ and $h_3\ll h_2$ where $[E_T]$ is the total enzyme concentration  \cite{thomas2011,*sanft2011}. The parameters in the reduced model are related to those in the developed model by $k_0=h_1$, $k_1=h_3[E_T]$, and $K=h_2/h_1$. This reduced master equation is solved exactly by a negative binomial distribution \cite{paulsson2000}.

The system size coefficients are obtained from Eq. (\ref{eqn:SSEcoeffs}), and are given by   
\begin{align}
 \label{enz:SSEcoeffs}
 \mathcal{D}_{n}^m= \delta_{m,0} k_0 + (-1)^n k_1 \frac{\partial^m}{\partial {[X]}^m}\frac{[X]}{K+[X]},
\end{align}
and $\mathcal{D}_{n,s}^{m}=0$ for $s>0$. In Fig. \ref{fig:NB}A and \ref{fig:NB}B, we consider two parameter sets corresponding to low and moderate numbers of substrate molecules, respectively. We observe that in contrast to the linear case, the discrete approximation of the nonlinear birth-death process tends to oscillate with increasing truncation order. This issue is addressed in the following section.

\section{Renormalization of nonlinear birth-death processes}

{
Van Kampen's ansatz, Eq. (\ref{eqn:ansatz}), bears the particularly simple interpretation that for linear birth-death processes $\epsilon$ denotes the fluctuations about the average given by the solution of the rate equation $[X]$. As noted in the previous example, for nonlinear birth-death processes these estimates are only approximate. Their asymptotic series expansions will therefore require additional terms that compensate for the deviations of the LNA from the true concentration mean and variance. It would therefore be desirable to find an approximation for nonlinear processes that yields more accurate mean and variance than the LNA. For instance by rewriting van Kampen's ansatz as
\begin{align}
 \label{eqn:ansatzRen}
 \frac{n}{\Omega}=\underbrace{\bigl.[X]+\Omega^{-1/2}\langle\epsilon\rangle\bigr.}_{\text{mean}} + \underbrace{\bigl.\Omega^{-1/2}\bar{\epsilon}\bigr.}_{\text{fluctuations}}.
\end{align}
Here, $\bar{\epsilon}=\epsilon-\langle\epsilon\rangle$ denotes a centered variable that quantifies the fluctuations about the true average which is a priori unknown. These estimates can however be approximated using the SSE beforehand, and the asymptotic expansion of the distributions can then be performed about these new estimates.
This idea is called renormalization and makes use of the fact that the terms correcting mean and variances can be summed exactly. As we show in the following the resummation allows to better control the convergence by effectively reducing the number of terms in the summation while at the same time it retains the accuracy of the expansion.}

The system size expansion of the moments, Eq. (\ref{eqn:MomentsFromCoeffs}), yields the following estimates for mean and variance of the fluctuations

\begin{subequations}
\label{eqn:sseMeanVar}
\begin{align}
\langle\epsilon\rangle &= \sum_{j=0}^N \Omega^{-j/2} a_1^{(j)}+ O(\Omega^{-(N+1)/2}),\\
{\bar{\sigma}^2}
  &= \sigma^2 + \sum_{j=1}^N \Omega^{-j/2} {\sigma^2_{(j)}}+ O(\Omega^{-(N+1)/2}),
\end{align}
\end{subequations}
respectively, where $\bar{\sigma}_{(j)}^2 = 2(a_2^{(j)}- {B_{j,2}(\{\chi!a_1^{(\chi)}\}_{\chi=1}^{j-1})}/{j!})$
and $B_{j,n}$ are the partial Bell polynomials \cite{comtet1974,*Note1}.

The renormalization procedure amounts to replacing $y$ by $\bar{y}=(n-\Omega[X]-\Omega^{1/2}\langle \epsilon \rangle)$, $\Sigma^2$ by $\bar{\Sigma}^2 = \Omega \bar{\sigma}^2$ in Eq. (\ref{eqn:P0})
and associating a new Gaussian $\bar{P}_0(n)$ with these estimates. The renormalized expansion is then given by

\begin{align}
 \label{eqn:RenormalizedExpansion}
 P(n,t) &= \bar{P}_0(n,t) \notag \\
       &+ \sum_{j=1}^{N} \Omega^{-j/2}\sum_{m=1}^{3j} \bar{a}_m^{(j)} \left(-\Omega^{1/2}\partial_n\right)^m  \bar{P}_0(n,t)
\nonumber\\& + O(\Omega^{-(N+1)/2}),
\end{align}
where the renormalized coefficients can be calculated from the bare ones using

\begin{align}
 \label{eqn:renormCoeffs}
 \bar{a}_m^{(j)}=\sum_{k=0}^{j}\sum_{n=0}^{3k} a_{n}^{(k)}\kappa_{m-n}^{(j-k)},
\end{align}
and 

\begin{align}
 \label{eqn:kappa}
 \kappa_j^{(n)} = 
 &\frac{1}{n!}\sum_{m=0}^{\lfloor j/2 \rfloor} (-1)^{(j+m)} \sum_{k=j-2m}^{n-m} { n \choose k}  \times
  \nonumber \\ &  B_{k,j-2m}\left(\left\{\chi!a_1^{(\chi)}\right\}\right) B_{n-k,m}\left(\left\{\frac{\chi!}{2}\bar{\sigma}_{(\chi)}^2\right\}\right),
\end{align}
where again $B_{k,n}(\{x_\chi\})$ denote the partial Bell polynomials \cite{Note1}. The result is verified at the end of this section. Note that the renormalized series has generally less non-zero coefficients since by construction $\bar{a}_1^{(j)}=\bar{a}_2^{(j)}=0$. Note that for linear birth-processes, mean and variance are exact to order $\Omega^0$ (LNA), and hence for this case expansion (\ref{eqn:GeneralizedExpansion}) coincides with Eq. (\ref{eqn:RenormalizedExpansion}).

For example, truncating after $\Omega^{-1}$-terms, from Eq. (\ref{eqn:sseMeanVar}) it follows that 
$\langle\epsilon\rangle = \Omega^{-1/2} a_1^{(1)}+O(\Omega^{-3/2})$ and $\bar{\sigma}^2=\sigma^2 + \Omega^{-1} (2 a_2^{(2)}-(a_1^{(1)})^2) + O(\Omega^{-3/2})$. Using Eq. (\ref{eqn:renormCoeffs}) the renormalized coefficients can be expressed in terms of the bare ones

\begin{subequations}
  \begin{align}
  &\bar{a}_1^{(1)} = 0,\ \ \bar{a}_3^{(1)} = a_3^{(1)}, \\
  &\bar{a}_2^{(2)}=0, \ \
  \bar{a}_4^{(2)}=a_{4}^{(2)}-a_{1}^{(1)} a_{3}^{(1)}, \ \
  \bar{a}_6^{(2)} = a_6^{(2)}.
  \end{align}
\end{subequations}

{This result can for instance be used to renormalize the stationary solution using the bare coefficients given in Sec. \ref{sec:steadystate}, Eqs. (\ref{eqn:O-1sol2}-\ref{eqn:O-1sol3}). The non-zero renormalized coefficients evaluate to
\begin{subequations}
\label{eqn:renormalizedCoeffSol}
\begin{align}
\bar{a}_3^{(1)} =& -\frac{\sigma^4 \mathcal{D}_1^{2}}{6\mathcal{J}}+\frac{\sigma^2 \mathcal{D}_2^{1}}{6\mathcal{J}}+\frac{\mathcal{D}_3^{0}}{18 \mathcal{J}},\\
\bar{a}_4^{(2)} =&
-\frac{\mathcal{D}_{4}^0}{96 \mathcal{J}}-\frac{\sigma ^2 \mathcal{D}_{3}^1}{24 \mathcal{J}}-\frac{\sigma ^4 \mathcal{D}_{2}^2}{16 \mathcal{J}}-\frac{\sigma^6 \mathcal{D}_{1}^3}{24 \mathcal{J}}\notag\\
&-\bar{a}_{3}^{(1)} \left(\frac{3 \mathcal{D}_{2}^1}{8 \mathcal{J}}+\frac{3 \sigma ^2 \mathcal{D}_{1}^2}{4 \mathcal{J}}\right),\notag\\
\bar{a}_6^{(2)} =& \frac{1}{2} (\bar{a}_3^{\text{(1)}})^2.
\end{align}
\end{subequations}
Note that for linear birth-death processes $\mathcal{D}_{n,s}^{m}=0$ for $s>0$ and $m>1$, and hence the above equations reduce to Eqs. (\ref{eqn:O-1sol2}-\ref{eqn:O-1sol3}).}

\subsection{The renormalized approximation performs well for nonlinear birth-death processes}

For the metabolic reaction (\ref{reaction:MM}), mean and variance can be obtained to be $\langle\epsilon\rangle = \Omega^{-1/2}\ess + O(\Omega^{-2})$, $\bar{\sigma}^2=\sigma^2+\Omega^{-1} \ess  (\ess +1) + O(\Omega^{-2})$, where $\ess=[X]/K$ is the reduced substrate concentration and $\sigma^2=K \ess  (\ess +1) $. Substituting now Eq. (\ref{enz:SSEcoeffs}) into Eqs. (\ref{eqn:renormalizedCoeffSol}), we obtain the expansion coefficients
\begin{subequations}
\label{enz:RenCoeffs1}
\begin{align}
 &\bar{a}_3^{\text{(1)}} = \frac{\sigma^2}{6} (2 \ess +1) , \\
 &\bar{a}_4^{\text{(2)}} = \frac{\sigma^2}{24} \left(6 \ess(\ess+1) +1\right), \ \
  \bar{a}_6^{\text{(2)}} = \frac{1}{2} (\bar{a}_3^{\text{(1)}})^2,
\end{align}
which determine the renormalized series expansion to order $\Omega^{-1}$. Using Eq. (\ref{eqn:coefficients}), (\ref{enz:SSEcoeffs}) and (\ref{eqn:renormCoeffs}) we can give the next order terms to order $\Omega^{-3/2}$ analytically  
\begin{align}
\label{enz:RenCoeffs2}
 &\bar{a}_3^{\text{(3)}} = \frac{\bar{a}_3^{\text{(1)}}}{K},\ \ 
 \bar{a}_5^{\text{(3)}} = \frac{\bar{a}_3^{\text{(1)}}}{20}(12 \ess (\ess +1) +1), \notag \\ 
 &\bar{a}_7^{\text{(3)}} = \bar{a}_3^{\text{(1)}}\bar{a}_4^{\text{(2)}}, \ \ 
 \bar{a}_9^{\text{(3)}} = \frac{1}{6} (\bar{a}_3^{\text{(1)}})^3.
\end{align}
\end{subequations}
In Fig. \ref{fig:NB}C and \ref{fig:NB}D we compare the renormalized approximation given by Eq. (\ref{eqn:RenormalizedExpansion}) with the respective bare approximations in Fig. \ref{fig:NB}A and \ref{fig:NB}B. We observe that the renormalization technique avoids oscillations and even the simple analytical approximation given by Eqs. (\ref{enz:RenCoeffs1}) is in reasonable agreement with the exact result. We note that the asymptotic approximations shown in C and D are almost indistinguishable for higher truncation orders.

\subsection{Proof of the renormalization formula}
   
The renormalized coefficients can in principle be obtained by matching the expansions given by Eq. (\ref{eqn:GeneralizedExpansion}) and (\ref{eqn:RenormalizedExpansion}) via their characteristic functions. For convenience we consider the characteristic function of the series (\ref{eqn:HermiteExpansion}) 

\begin{align}
 \label{eqn:CF}
 G(k) = G_0(k) \left( 1+ \sum_{j=1}^\infty \Omega^{-j/2} \sum_{n=1}^{3j} a_n^{(j)} (ik)^n  \right),
\end{align}
with $G_0(k)=e^{-(k\sigma)^2/2}$ being the characteristic function solution of the LNA $\pi_0(\epsilon)$ and we have omitted the explicit time-dependence to ease the notation. We are now looking for a different expansion with corrected estimates for the mean  and variance.

\begin{align}
 \label{eqn:CFR}
 \bar{G}(k) 
&=\bar{G}_0(k) \left( 1+ \sum_{j=1}^\infty \Omega^{-j/2} \sum_{n=1}^{3j} \bar{a}_n^{(j)} (ik)^n  \right),
\end{align}
Note that $\bar{G}_0(k)=e^{ik\langle\epsilon\rangle}e^{-(k\bar{\sigma})^2/2}$ is the characteristic function for a Gaussian random variable with mean $\langle\epsilon\rangle$ and variance $\bar{\sigma}^2$ given by Eqs. (\ref{eqn:sseMeanVar}). 

Equating now Eq. (\ref{eqn:CF}) and (\ref{eqn:CFR}), we find

\begin{align}
 \label{eqn:deee}
 1+& \sum_{j=1}^\infty \Omega^{-j/2} \sum_{n=1}^{3j} \bar{a}_n^{(j)} (ik)^n \nonumber\\
 &= \frac{G_0(k)}{\bar{G}_0(k)}
 \left( 1+ \sum_{j=1}^\infty \Omega^{-j/2} \sum_{n=1}^{3j} a_n^{(j)} (ik)^n  \right).
\end{align}
Expanding the prefactor in the above equation in powers of $k$ and then in $\Omega$, we have

\begin{align}
\label{eqn:prefactor}
\frac{G_0(k)}{\bar{G}_0(k)}
=
\sum_{j=0}^{\infty} (ik)^j \kappa_j=\sum_{n=0}^{\infty} \Omega^{-n/2} \sum_{j=0}^{2n} (ik)^j \kappa_j^{(n)},
\end{align}
from which Eq. (\ref{eqn:renormCoeffs}) follows, which expresses the new coefficients $\bar{a}_n^{(j)}$ in terms of the bare ones ${a}_n^{(j)}$. It remains to derive an explicit expression for the $\kappa_j^{(n)}$. The expansion in powers of $(ik)$ yields

\begin{equation}
 \label{eqn:pre_kappa}
 \kappa_j=\sum_{m=0}^{\lfloor j/2 \rfloor} (-1)^{(j+m)}
 \frac{{\langle \epsilon \rangle }^{j-2m}}{(j-2m)!}
 \frac{\left( \frac{\bar{\sigma}^2-\sigma^2}{2}\right)^{m}}{m!}. 
\end{equation}
We now expand the first term in inverse powers of $\Omega$ using the partial Bell polynomials 

\begin{align}
 \frac{1}{(j-2m)!} &  \left( \sum_{n=1}^\infty \Omega^{-n/2} a_1^{(n)} \right)^{j-2m} \notag\\
 =\sum_{n=1}^\infty & \frac{\Omega^{-n/2}}{n!}  \sum_{k=0}^n \delta_{j-2m,k} B_{n,k} \left(\left\{  \chi! a_1^{(\chi)}   \right\}\right),
\end{align}
and similarly for the second term

\begin{align}
 \frac{1}{m!} &\left( \frac{1}{2}\sum_{n=1}^\infty \Omega^{-n/2} {\bar{\sigma}_{(n)}^2} \right)^{m} \nonumber \\
 &= \sum_{n=0}^\infty \frac{\Omega^{-n/2}}{n!} \sum_{k=0}^n \delta_{m,k} B_{n,k}\left(\left\{\frac{\chi!}{2}\bar{\sigma}_{(\chi)}^2\right\}\right).
\end{align}
Using the above expansions in Eq. (\ref{eqn:pre_kappa}) and rearranging in powers of $\Omega^{-1/2}$, Eq. (\ref{eqn:kappa}) for the coefficients $\kappa_j^{(n)}$ follows.

Finally, one associates with the centered variable $\bar{\epsilon}=\epsilon-\langle\epsilon\rangle$, a Gaussian $\bar{\pi}_0(\bar{\epsilon})$ with variance $\bar{\sigma}^2$. It then follows from inverting Eq. (\ref{eqn:CFR}) that
$
 \Pi(\bar{\epsilon})
= \bar{\pi}_0(\bar{\epsilon}) + \sum_{j=1}^N \Omega^{-j/2} \sum_{n=1}^{3j} \bar{a}_n^{(j)} \psi_n\left(\bar{\epsilon}\right)  \bar{\pi}_0(\bar{\epsilon})+O(\Omega^{-(N+1)/2})
$. Associating now the $\Omega^{-j/2}$-term of this equation with $\pi_j$ in Eq. (\ref{eqn:connectionformula}), the discrete series for $P(n,t)$ given by Eq. (\ref{eqn:RenormalizedExpansion}) follows.

\section{Application}
\label{sec:application}

\begin{figure}
  \includegraphics[width=0.45\textwidth]{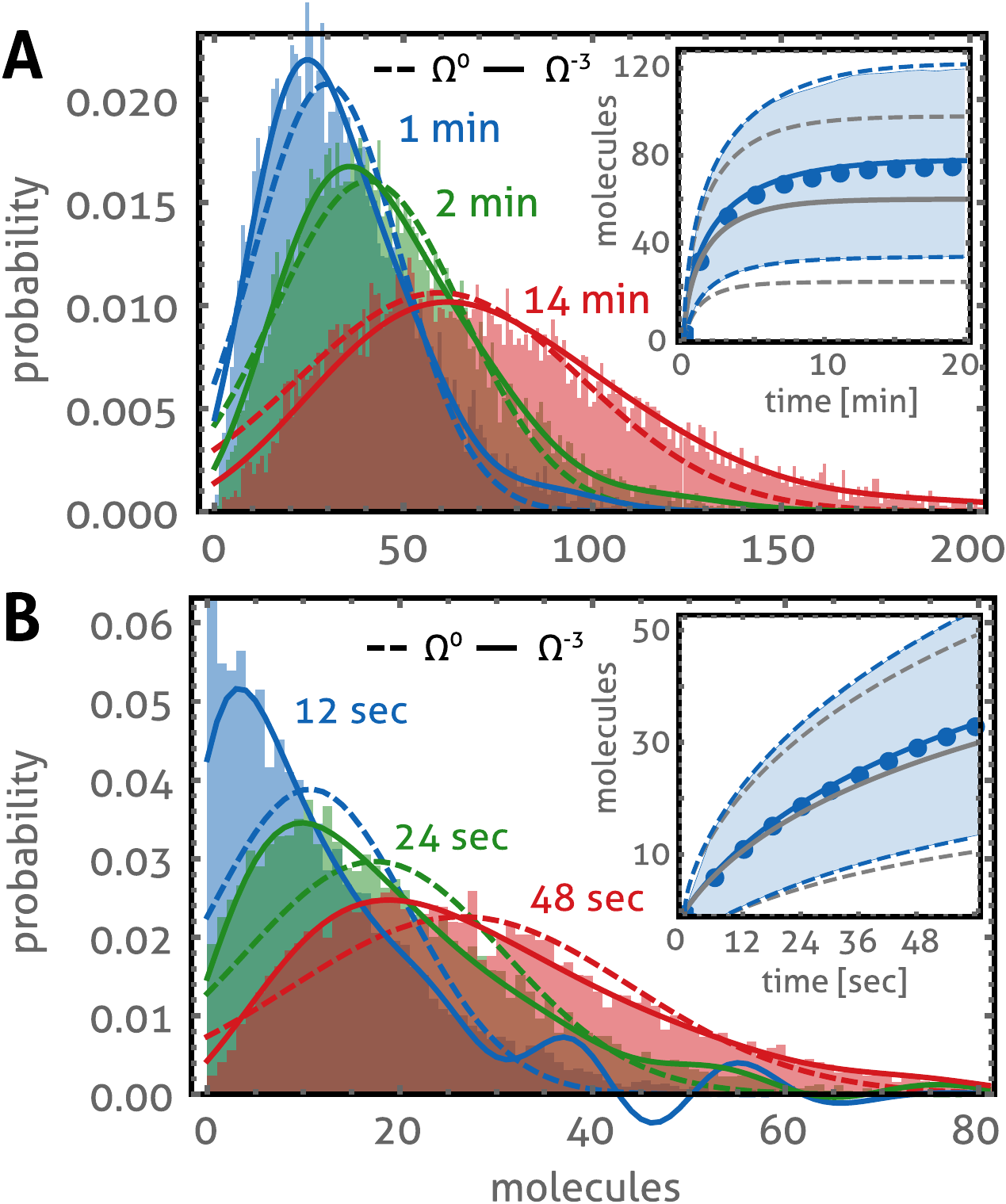}
   \caption{(Color
 online) \textbf{Predicting transient distributions of gene expression.} The dynamics of protein synthesis with enzymatic degradation, scheme (\ref{reaction:gene}), is studied using the burst approximation (\ref{appl:burstapprox}). 
   (A) We compare the time-dependence of the renormalized discrete approximations to exact stochastic simulations at times $1$, $2$, and $14 min$. 
   The overall shape (mode, skewness, distribution tails) of the simulated distributions (bars) is in excellent agreement with the series approximation when truncated after $\Omega^{-3}$-terms (solid lines) but not when only $\Omega^{0}$ are taken into account (dashed lines). This agreement is also observed for the first two moments shown in the inset: while the $\Omega^{-3}$-approximation (blue solid line) agrees with the moment dynamics of the simulated distributions (dots) of the reduced model (\ref{appl:burstapprox}), the $\Omega^{0}$-approximation underestimates the mean (gray solid line) and variance by $25\%$. The area within one standard deviation of the mean obtained from simulations is shown in blue, the boundary obtained from the approximations are shown as dashed lines ($\Omega^{0}$ grey, $\Omega^{-3}$ blue).
   (B) Despite the good agreement shown in (A) we found that at very short times ($12s$ -- blue solid line) the series truncated after $\Omega^{-3}$-terms tends to oscillations which quickly disappear for later times ($24s$ -- green, $48 s$ -- red solid line). See main text for discussion. Parameters are $k_0\Omega = 15 {min}^{-1}$, $k_1\Omega=100 {min}^{-1}$, $K\Omega = 20$, $\Omega=100$, and $b = 5$. Histograms were obtained from $10,000$ stochastic simulations.
   }
  \label{fig:Fig3}
\end{figure}

The models studied so far have been useful to develop the method. It remains however to be demonstrated that it remains accurate in cases where analytical solution is not feasible, as for instance, for out-of-steady-state and non-detailed balance systems. We here consider the synthesis of a protein $P$ which is degraded through an enzyme

\begin{subequations}
\label{reaction:gene}
\begin{align}
 &\varnothing \xrightarrow[]{h_0} M \xrightarrow[]{h_1} \varnothing,  \ \
 M \xrightarrow[]{h_2} M + P, \label{appl:synthesis} \\
 &P + E \xrightleftharpoons[h_4]{h_3} C \xrightarrow[]{h_5} E, \label{appl:degr}
\end{align} 
\end{subequations}
where $M$ denotes the transcript, $E$ the  enzyme and $C$ complex species as has been studied in Ref. \cite{thomas2012}. Since our theory applies only to a single species, we consider the limiting case in which the protein dynamics represents the slowest timescale of the system. It has be shown \cite{shahrezaei2008} that when species $M$ is degraded much faster than the protein $P$, the protein synthesis (\ref{appl:synthesis}) reduces to the transition $S_1=+z$, $\gamma_1=\Omega k_0$ in which $z$ is a random variable following the geometric distribution $\varphi(z)= \frac{1}{1+b}\left(\frac{b}{1+b}\right)^z$ with average $b$, which is called the burst approximation. Similarly to the metabolic reaction studied in Sec. \ref{sec:discrete}, the enzymatic degradation process (\ref{appl:degr}) can be reduced to $S_2=-1$, $\gamma_2=\Omega k_1 \frac{n}{\Omega K+n}$ with a nonlinear dependence on the protein number $n$. The master equation describing the protein number is then given by

\begin{align}
 \label{appl:burstapprox}
 \frac{\mathrm{d}}{\mathrm{d}t} P(n) =&  \Omega  \sum_{z=0}^\infty (E^{-z}-1) k_0 \varphi(z) P(n) \notag \\
                    &+ \Omega (E^{+1}-1)  k_1 \frac{n}{\Omega K+n} P(n).
\end{align}
The relation between the parameters in the reduced and the developed model are given by $k_0=h_0 h_2 / h_1$, $b=h_2/h_1$, $k_1=h_5 [E_T]$, $K=h_5/h_3$, where $[E_T]$ denotes the total enzyme concentration. This description involves countably many reactions: one for the degradation of the protein, and one for each value of $z$. Therefore, the reactions cannot obey detailed balance in steady state. The system size coefficients now follow from Eq. (\ref{eqn:SSEcoeffs}), and are given by   

\begin{align}
 \label{appl:SSEcoeffs}
 \mathcal{D}_{n}^{m}= \delta_{m,0} k_0 \langle z^n \rangle_\varphi + (-1)^n k_1 \frac{\partial^m}{\partial {[X]}^m}\frac{[X]}{K+[X]},
\end{align}
and $\mathcal{D}_{n,s}^{m}=0$ for $s>0$, where $\langle z^n \rangle_\varphi=\sum_{z=0}^\infty z^n\varphi(z)=\frac{1}{1+b}\operatorname{Li}_{-n}(\frac{b}{1+b})$ denotes the average over the geometric distribution in terms of the polylogarithm function \cite{Note2}. The deterministic equation is given by

\begin{align}
 \label{appl:REs}
 \frac{\mathrm{d} [X]}{\mathrm{d} t} = k_0 b - \frac{k_1[X]}{K+[X]},
\end{align}
which follows from the expression for $\mathcal{D}_1^0$. Using the Jacobian $\mathcal{J}=\mathcal{D}_1^1$ and $\mathcal{D}_2^0$ in Eq. (\ref{eqn:lyapunov}), we find that the LNA variance obeys

\begin{align}
 \label{appl:LNA}
 \frac{\partial \sigma^2}{\partial t}=-\frac{2 k_1 K}{([X]+K)^2} \sigma^2 + k_0 b(1+2b) +\frac{ k_{1} [X]}{K+[X]}.
\end{align}
The ODEs given by Eq. (\ref{appl:REs}) and (\ref{appl:LNA}) are integrated numerically and the solution is used in Eq. (\ref{eqn:P0}) from which the leading order approximation follows. Higher order approximations are now be obtained by using Eq. (\ref{appl:SSEcoeffs}) in (\ref{eqn:coefficientEquations}) which govern the time-evolution of the coefficients $a_m^{(j)}(t)$ and using the result in Eq.  (\ref{eqn:RenormalizedExpansion}) and (\ref{eqn:renormCoeffs}). We assume deterministic initial conditions with zero proteins meaning 
$a_m^{(j)}(0)=\delta_{m,0}\delta_{j,0}$. In Fig. \ref{fig:Fig3}A we compare the time-evolution obtained by the leading order approximation $P_0$ and Eq. (\ref{eqn:RenormalizedExpansion}) truncated after the $\Omega^{-3}$-term. The latter distributions are in excellent agreement with the distributions sampled using the stochastic simulation algorithm \cite{gillespie1977,*Note3}. In particular, unlike the leading order approximation, these describe well mode, skewness, and tails of the distribution. We note that also the mean and variance of these distribution approximations are in excellent agreement as verified in inset of Fig. \ref{fig:Fig3}A.

Despite the overall good agreement, in Fig. \ref{fig:Fig3}B we show that there are discrepancies at very short times where and, again, the distribution approximations tend to oscillations. Motivated by this numerical observation, we speculate that this behavior of the expansion is due a temporal boundary layer as commonly observed in singular perturbation expansions \cite{verhulst2006}. Theoretically, the layer must be located at times of the same order as the expansion parameter, i.e., $t=(\Omega K)^{-1/2}min\approx 13 s$, coinciding with the simulation in Fig. \ref{fig:Fig3}B. This suggests that our approach does only describe the outer solution. Further analysis would be required to investigate also the inner solution which is beyond the scope of this article.

\section{Discussion}

We have here presented an approximate solution method for the probability distribution of the master equation. The solution is given in terms of an asymptotic series expansion that can be truncated systematically to any desired order in the inverse system size. For biochemical systems with large numbers of molecules, we have derived a continuous series approximation that extends van Kampen's LNA to higher orders in the SSE. In low molecule number conditions, we have found that this continuous approximation becomes inaccurate. Instead, in most practical situations the prescribed discrete distribution approximations incorporating higher order terms in the SSE better capture the underlying solution of the master equation. {While the terms to order $\Omega^{-1}$ have been given explicitly, we found that for the examples studied here up to $\Omega^{-3}$ or $\Omega^{-4}$-terms had to be taken into account
to accurately characterize these non-Gaussian distributions. We note, however, that the asymptotic expansion cannot generally guarantee the positivity of the probability law. These undulations are particularly pronounced in the short-time behavior of the expansion studied in Sec. \ref{sec:application}, which our theory does not describe.}

{Previous means of solving the master equation have either been numerical in nature \cite{munsky2006} or have focused on the inverse problem, i.e., reconstruction of the probability density from the moments.
While a numerical solution for the master equation of a single species is rather straightforward, we expect our procedure to become computationally advantageous when generalized to the multivariate case where numerical solution is usually prohibitive because of combinatorial explosion.}

{
Methods based on moments typically require approximations such as moment closure \cite{smadbeck2013,*alexander2014} and also require the prior assumption of the first few moments containing all information on the probability distribution. Conversely, using the system size expansion, we have here obtained the probability distribution directly from the master equation \emph{without the need to resort to moments}. 
This method enjoys the particular advantage over previous ones that the first few terms of this expansion can be written down explicitly as a function of the rate constants and for any number of reactions. For small models we have demonstrated that the procedure leads to particularly simple expressions for the non-Gaussian distributions. 
This development could prove particularly valuable for parameter estimation of biochemical reactions in living cells.}

\begin{acknowledgments}
 It is a pleasure to thank Claudia Cianci and David Schnoerr for careful reading of the manuscript.
\end{acknowledgments}

\setcounter{equation}{0}
\renewcommand{\theequation}{A\arabic{equation}}

\appendix

\section{Useful properties of the Hermite polynomials}
\label{app:Hermite}

We here briefly review some properties of the Hermite orthogonal polynomials. The polynomials can be defined in terms of the derivatives of a centered Gaussian $\pi_0$ with variance $\sigma^2$,

\begin{align}
 H_n \left(\frac{\epsilon}{\sigma}\right) = \pi_0^{-1}(\epsilon) (-\sigma\partial_\epsilon)^{n} \pi_0(\epsilon).
\end{align}
An explicit formula is

\begin{align}
 \label{eqn:ExplicitHermite}
 H_n\left(\frac{\epsilon}{\sigma}\right)
= \sum_{k=0}^{\lfloor n/2\rfloor} {n \choose 2k} (-1)^k\left({\epsilon\over \sigma}\right)^{n-2k} (2k-1)!!\,.
\end{align}
These functions are orthogonal
$ \frac{1}{n!}\int_{-\infty}^\infty \mathrm{d}\epsilon\, {\mathit{H}}_m\left(\frac{\epsilon}{\sigma}\right) {\mathit{H}}_n\left(\frac{\epsilon}{\sigma}\right)\,  \pi_0(\epsilon) =$ $\delta_{nm}$,
with respect to the Gaussian measure $\pi_0$. The derivative satisfies 

\begin{align}
 \label{eqn:HermiteDerivative}
 (\sigma\partial_\epsilon)^{m} H_n\left(\frac{\epsilon}{\sigma}\right)=\frac{n!}{(n-m)!} H_{n-m}\left(\frac{\epsilon}{\sigma}\right).
\end{align}
Since these polynomials are complete, every function $f(\epsilon)$ in $L_2(\mathbb{R},\pi_0)$ (not necessarily positive) can be expanded as
$
 f(\epsilon) = \sum_{n=0}^\infty b_n H_n\left(\frac{\epsilon}{\sigma}\right) \pi_0(\epsilon),
$
where the coefficients are given by $b_n=\frac{1}{n!}\int \mathrm{d}\epsilon\, H_n\left(\frac{\epsilon}{\sigma}\right)f(\epsilon)$.
We note because $H_0\left(\frac{\epsilon}{\sigma}\right)=1$ and $\pi_0$ is normalized, we must have $b_0=1$ if $\int \mathrm{d} \epsilon\, f(\epsilon)=1$.

\section{Explicit derivation of Eq. (\ref{eqn:integral}) and the properties of the expansion coefficients}
\label{app:integral}


Changing variables $\epsilon=x\sigma$ and letting $\tilde{\mathcal{I}}_{mn}^{\alpha\beta} = \sigma^{\alpha-\beta+m-n} {\mathcal{I}}_{mn}^{\alpha\beta}$, the integral (\ref{eqn:integraldef}) can be written
\begin{equation}
\label{app:integraldef}
\tilde{\mathcal{I}}^{\alpha\beta}_{mn} = \frac{1}{n!\alpha!\beta!} \int \mathrm{d}x H_n\left(x\right) (-\partial_x)^\alpha x^\beta H_m\left(x\right) \pi_0(x),
\end{equation}
where $\pi_0(x)$ is a centered Gaussian with unit variance.
Using partial integration, property (\ref{eqn:HermiteDerivative}), and the relation

\begin{align}
 H_\alpha(x) H_\beta(x) = \alpha!\beta! \sum_{s=0}^{\min(\alpha,\beta)} \frac{H_{\alpha+\beta-2s}(x)}{s!(\alpha-s)!(\beta-s)!},
\end{align}
given in Ref. \cite{Lebedev}, one obtains

\begin{align}
 \tilde{\mathcal{I}}^{\alpha \beta}_{mn}
  &= \frac{1}{\alpha!\beta!}\sum_{s=0}^{\min({n-\alpha,m})}\binom ms \frac{\int \mathrm{d} x\,  x^\beta H_{m+n-\alpha-2s}(x) \pi_0(x)}{(n-\alpha-s)!}.
\end{align}
The remaining integral can now be evaluated as the moments of the unit Gaussian which yields

\begin{align}
\label{app:HermiteMoments}
\int \mathrm{d} x\, x^b H_a(x) \pi_0(x) 
&= \frac{b!}{(b-a)!}\int\mathrm{d} x\, x^{b-a} \pi_0(x) \nonumber\\
&= \frac{b!}{(b-a)!}(b-a-1)!!\,.
\end{align}
for even $(b-a) \ge 0$ and zero otherwise. Explicitly, the matrix elements are given by

\begin{align}
 \tilde{\mathcal{I}}^{\alpha\beta}_{mn} 
   =&  \frac{1}{\alpha!}\sum_{s=0}^{\min({n-\alpha,m})}\binom ms \times \nonumber \\
& \frac{(\beta+\alpha+2s-(m+n)-1)!!}{(\beta+\alpha+2s-(m+n))!(n-\alpha-s)!},
\end{align}
for even $(\alpha+\beta)-(m+n)$ and zero otherwise. Note that the above quantity is strictly positive.
Note also that the argument of the double factorial is taken to be positive and hence the summation is non-zero only if
$
\alpha+\beta+2\min({n-\alpha,m})\ge m+n
$
and hence for even $\beta=2k$ we have
$
n=m+\alpha\pm 2l,
$
while for odd $\beta=(2k+1)$ we have
$n=m+\alpha\pm (2l+1)$, with $l=0,\ldots, k$.

{
The integral formula can be used to verify two important properties of the solution of Eq. (\ref{eqn:coefficientEquations}) given deterministic initial conditions:
(i)
We have $N_j=3j$ and hence Eq. (\ref{eqn:coefficientEquations}) indeed yields a finite number of equations.
(ii)
The coefficients $a_n^{(j)}$ for which $(n+j)$ is odd vanish at all times.}

{To verify property (i), let $N_j$ be the index of the highest eigenfunction required to order $\Omega^{-j/2}$. Using Eq. (\ref{eqn:coefficientEquations}) one can show that $a_{N_j}^{(j)} \sim a_{N_{j-1}}^{(j-1)} \mathcal{I}_{N_{j-1},N_j}^{p,3-p}$ for $p \in \{1,2,3\}$. By virtue of the properties given after Eq. (\ref{eqn:integral}), we find $N_j=N_{j-1}+3$. Since for deterministic initial conditions we have $N_0=0$, it follows that $N_j=3j$.}

{Finally, we verify property (ii). To the summation in Eq. (\ref{eqn:coefficientEquations}) there contribute only terms for which $ \mathcal{I}_{mn}^{p,k-p-2(s-1)}$ is non-zero. Hence, by the condition given after Eq. (\ref{eqn:integral}), $k-(m+n)$ is an even number. Considering the equation for $a_n^{(j)}$ for which $n+j$ is even, it follows that in the summation on the right hand side of Eq. (\ref{eqn:coefficientEquations}) there appear only coefficients for which $m+(j-k)$ is even. Conversely, for $n+j$ being odd then same holds for $m+(j-k)$. Hence the pairs of equations for $a_n^{(j)}$ for which $(j+n)$ is even or odd are mutually uncoupled. For deterministic initial conditions, only terms with $j+n$ even differ from zero initially from which the result follows.}

\section{Solution to the problem of moments using the system size expansion}
\label{app:problemofmoments}

%
%
 Having obtained the moment expansion in terms of the coefficients $a_n^{(j)}$, it would be desirable to invert this relation and the coefficients in terms of the expansion of the moments. This can be derived using the completeness of the Hermite polynomials, and writing the probability density as
$
 \Pi(\epsilon) =  \sum_{n=0}^\infty b_n {H_n\left(\frac{\epsilon}{\sigma}\right)}  \pi_0(\epsilon),
$
where the $b_n= \frac{1}{n!}\int d\epsilon H_n\left(\frac{\epsilon}{\sigma}\right) \Pi(\epsilon)$ can be expressed in terms of the moments using Eq. (\ref{eqn:ExplicitHermite}), as follows

\begin{align}
 b_n %
 &= \frac{1}{n!} \sum_{k=0}^{\lfloor n/2\rfloor} {n \choose 2k} (-1)^{k} \frac{\langle \epsilon^{n-2k} \rangle}{\sigma^{n-2k}} (2k-1)!!.
\end{align}
Assuming now that the moments can be expanded in a series in powers of $\Omega$, i.e.,

\begin{align}
 \label{app:SSEmoments}
 \langle \epsilon^\beta \rangle = \sum_{j=0}^N \Omega^{-j/2} [\epsilon^\beta]_j + O(\Omega^{-(N+1)/2}),
\end{align}
the $b_n$ can be matched to the coefficients $a_n$ in Eq. (\ref{eqn:HermiteExpansion}) using
$ \sigma^n b_n = \sum_{j=0}^{N} {\Omega^{-j/2}} a_n^{(j)} + O(\Omega^{-(N+1)/2})$, from which one obtains

\begin{align}
 \label{eqn:CoeffFromMoments}
 a_n^{(j)}=\frac{1}{n!} \sum_{k=0}^{\lfloor n/2\rfloor} {n \choose 2k} (-\sigma^2)^k (2k-1)!! [ \epsilon^{n-2k} ]_j \,,
\end{align}
with $[\epsilon^0]_j=\delta_{j,0}$. The above formula relates the expansion of the moments to the expansion of distribution functions. It is now evident that the system size expansion of the distribution can be constructed from the system size expansion for a finite set of moments. 
%
%

Specifically, to order $\Omega^{-1/2}$ the non-zero coefficients evaluate to
\begin{align}
 a_1^{(1)}=[\epsilon]_1, \ \
 a_3^{(1)}=\frac{1}{3!}\left([\epsilon^3]_1^3-3 \sigma ^2 [\epsilon]_1\right)
\end{align}
while the coefficients to order $\Omega^{-1}$ are given by
\begin{align}
a_2^{(2)}&=
\frac{1}{2} [\epsilon^2]_2, \ \
a_4^{(2)}=\frac{1}{4!} \left(\text{[$\epsilon^4$]}_2-6 \sigma ^2 \text{[$\epsilon^2$]}_2\right), \notag\\
a_6^{(2)}&=\frac{1}{6!} \left(45 \sigma ^4 \text{[$\epsilon^2 $]}_2-15 \sigma ^2 \text{[$\epsilon^4 $]}_2+\text{[$\epsilon^6 $]}_2\right).
\end{align}
A different series is obtained using the Edgeworth expansion which instead of using the system size expansion of the moments, Eq. (\ref{app:SSEmoments}), proceeds by scaling the cumulants by a size parameter.
%
%
%
%
%

\clearpage


\begin{thebibliography}{43}%
\makeatletter
\providecommand \@ifxundefined [1]{%
 \@ifx{#1\undefined}
}%
\providecommand \@ifnum [1]{%
 \ifnum #1\expandafter \@firstoftwo
 \else \expandafter \@secondoftwo
 \fi
}%
\providecommand \@ifx [1]{%
 \ifx #1\expandafter \@firstoftwo
 \else \expandafter \@secondoftwo
 \fi
}%
\providecommand \natexlab [1]{#1}%
\providecommand \enquote  [1]{``#1''}%
\providecommand \bibnamefont  [1]{#1}%
\providecommand \bibfnamefont [1]{#1}%
\providecommand \citenamefont [1]{#1}%
\providecommand \href@noop [0]{\@secondoftwo}%
\providecommand \href [0]{\begingroup \@sanitize@url \@href}%
\providecommand \@href[1]{\@@startlink{#1}\@@href}%
\providecommand \@@href[1]{\endgroup#1\@@endlink}%
\providecommand \@sanitize@url [0]{\catcode `\\12\catcode `\$12\catcode
  `\&12\catcode `\#12\catcode `\^12\catcode `\_12\catcode `\%12\relax}%
\providecommand \@@startlink[1]{}%
\providecommand \@@endlink[0]{}%
\providecommand \url  [0]{\begingroup\@sanitize@url \@url }%
\providecommand \@url [1]{\endgroup\@href {#1}{\urlprefix }}%
\providecommand \urlprefix  [0]{URL }%
\providecommand \Eprint [0]{\href }%
\providecommand \doibase [0]{http://dx.doi.org/}%
\providecommand \selectlanguage [0]{\@gobble}%
\providecommand \bibinfo  [0]{\@secondoftwo}%
\providecommand \bibfield  [0]{\@secondoftwo}%
\providecommand \translation [1]{[#1]}%
\providecommand \BibitemOpen [0]{}%
\providecommand \bibitemStop [0]{}%
\providecommand \bibitemNoStop [0]{.\EOS\space}%
\providecommand \EOS [0]{\spacefactor3000\relax}%
\providecommand \BibitemShut  [1]{\csname bibitem#1\endcsname}%
\let\auto@bib@innerbib\@empty
\bibitem [{\citenamefont {Jahnke}\ and\ \citenamefont
  {Huisinga}(2007)}]{jahnke2007}%
  \BibitemOpen
  \bibfield  {author} {\bibinfo {author} {\bibfnamefont {T.}~\bibnamefont
  {Jahnke}}\ and\ \bibinfo {author} {\bibfnamefont {W.}~\bibnamefont
  {Huisinga}},\ }\href@noop {} {\bibfield  {journal} {\bibinfo  {journal} {J
  Math Biol}\ }\textbf {\bibinfo {volume} {54}},\ \bibinfo {pages} {1}
  (\bibinfo {year} {2007})}\BibitemShut {NoStop}%
\bibitem [{\citenamefont {Haken}(1974)}]{haken1974}%
  \BibitemOpen
  \bibfield  {author} {\bibinfo {author} {\bibfnamefont {H.}~\bibnamefont
  {Haken}},\ }\href@noop {} {\bibfield  {journal} {\bibinfo  {journal} {Phys
  Lett A}\ }\textbf {\bibinfo {volume} {46}},\ \bibinfo {pages} {443} (\bibinfo
  {year} {1974})}\BibitemShut {NoStop}%
\bibitem [{\citenamefont {Van~Kampen}(1976{\natexlab{a}})}]{vanKampen1976}%
  \BibitemOpen
  \bibfield  {author} {\bibinfo {author} {\bibfnamefont {N.~G.}\ \bibnamefont
  {Van~Kampen}},\ }\href@noop {} {\bibfield  {journal} {\bibinfo  {journal}
  {Phys Lett A}\ }\textbf {\bibinfo {volume} {59}},\ \bibinfo {pages} {333}
  (\bibinfo {year} {1976}{\natexlab{a}})}\BibitemShut {NoStop}%
\bibitem [{\citenamefont {Mazo}(1975)}]{mazo1975}%
  \BibitemOpen
  \bibfield  {author} {\bibinfo {author} {\bibfnamefont {R.~M.}\ \bibnamefont
  {Mazo}},\ }\href@noop {} {\bibfield  {journal} {\bibinfo  {journal} {J Chem
  Phys}\ }\textbf {\bibinfo {volume} {62}},\ \bibinfo {pages} {4244} (\bibinfo
  {year} {1975})}\BibitemShut {NoStop}%
\bibitem [{\citenamefont {Peccoud}\ and\ \citenamefont
  {Ycart}(1995)}]{peccoud1995}%
  \BibitemOpen
  \bibfield  {author} {\bibinfo {author} {\bibfnamefont {J.}~\bibnamefont
  {Peccoud}}\ and\ \bibinfo {author} {\bibfnamefont {B.}~\bibnamefont
  {Ycart}},\ }\href@noop {} {\bibfield  {journal} {\bibinfo  {journal} {Theor
  Popul Biol}\ }\textbf {\bibinfo {volume} {48}},\ \bibinfo {pages} {222}
  (\bibinfo {year} {1995})}\BibitemShut {NoStop}%
\bibitem [{\citenamefont {Bokes}\ \emph {et~al.}(2012)\citenamefont {Bokes},
  \citenamefont {King}, \citenamefont {Wood},\ and\ \citenamefont
  {Loose}}]{bokes2012}%
  \BibitemOpen
  \bibfield  {author} {\bibinfo {author} {\bibfnamefont {P.}~\bibnamefont
  {Bokes}}, \bibinfo {author} {\bibfnamefont {J.~R.}\ \bibnamefont {King}},
  \bibinfo {author} {\bibfnamefont {A.~T.}\ \bibnamefont {Wood}}, \ and\
  \bibinfo {author} {\bibfnamefont {M.}~\bibnamefont {Loose}},\ }\href@noop {}
  {\bibfield  {journal} {\bibinfo  {journal} {J Math Biol}\ }\textbf {\bibinfo
  {volume} {64}},\ \bibinfo {pages} {829} (\bibinfo {year} {2012})}\BibitemShut
  {NoStop}%
\bibitem [{\citenamefont {G{\"o}rtz}\ and\ \citenamefont
  {Walls}(1976)}]{gortz1976}%
  \BibitemOpen
  \bibfield  {author} {\bibinfo {author} {\bibfnamefont {R.}~\bibnamefont
  {G{\"o}rtz}}\ and\ \bibinfo {author} {\bibfnamefont {D.}~\bibnamefont
  {Walls}},\ }\href@noop {} {\bibfield  {journal} {\bibinfo  {journal} {Z Phys
  B Con Mat}\ }\textbf {\bibinfo {volume} {25}},\ \bibinfo {pages} {423}
  (\bibinfo {year} {1976})}\BibitemShut {NoStop}%
\bibitem [{\citenamefont {Haag}\ and\ \citenamefont
  {H{\"a}nggi}(1979)}]{haag1979}%
  \BibitemOpen
  \bibfield  {author} {\bibinfo {author} {\bibfnamefont {G.}~\bibnamefont
  {Haag}}\ and\ \bibinfo {author} {\bibfnamefont {P.}~\bibnamefont
  {H{\"a}nggi}},\ }\href@noop {} {\bibfield  {journal} {\bibinfo  {journal} {Z
  Phys B Con Mat}\ }\textbf {\bibinfo {volume} {34}},\ \bibinfo {pages} {411}
  (\bibinfo {year} {1979})}\BibitemShut {NoStop}%
\bibitem [{\citenamefont {Schnoerr}\ \emph {et~al.}(2014)\citenamefont
  {Schnoerr}, \citenamefont {Sanguinetti},\ and\ \citenamefont
  {Grima}}]{schnoerr2014}%
  \BibitemOpen
  \bibfield  {author} {\bibinfo {author} {\bibfnamefont {D.}~\bibnamefont
  {Schnoerr}}, \bibinfo {author} {\bibfnamefont {G.}~\bibnamefont
  {Sanguinetti}}, \ and\ \bibinfo {author} {\bibfnamefont {R.}~\bibnamefont
  {Grima}},\ }\href@noop {} {\bibfield  {journal} {\bibinfo  {journal} {J Chem
  Phys}\ }\textbf {\bibinfo {volume} {141}},\ \bibinfo {pages} {024103}
  (\bibinfo {year} {2014})}\BibitemShut {NoStop}%
\bibitem [{\citenamefont {Van~Kampen}(1976{\natexlab{b}})}]{van1976}%
  \BibitemOpen
  \bibfield  {author} {\bibinfo {author} {\bibfnamefont {N.~G.}\ \bibnamefont
  {Van~Kampen}},\ }\href@noop {} {\bibfield  {journal} {\bibinfo  {journal}
  {Adv Chem Phys}\ }\textbf {\bibinfo {volume} {34}},\ \bibinfo {pages} {245}
  (\bibinfo {year} {1976}{\natexlab{b}})}\BibitemShut {NoStop}%
\bibitem [{\citenamefont {Van~Kampen}(1997)}]{vanKampen}%
  \BibitemOpen
  \bibfield  {author} {\bibinfo {author} {\bibfnamefont {N.~G.}\ \bibnamefont
  {Van~Kampen}},\ }\href@noop {} {\emph {\bibinfo {title} {Stochastic Processes
  in Physics and Chemistry.}}},\ \bibinfo {edition} {3rd}\ ed.\ (\bibinfo
  {publisher} {Elsevier, Amsterdam},\ \bibinfo {year} {1997})\BibitemShut
  {NoStop}%
\bibitem [{\citenamefont {Elf}\ and\ \citenamefont
  {Ehrenberg}(2003)}]{ElfEhrenberg}%
  \BibitemOpen
  \bibfield  {author} {\bibinfo {author} {\bibfnamefont {J.}~\bibnamefont
  {Elf}}\ and\ \bibinfo {author} {\bibfnamefont {M.}~\bibnamefont
  {Ehrenberg}},\ }\href@noop {} {\bibfield  {journal} {\bibinfo  {journal}
  {Genome Res}\ }\textbf {\bibinfo {volume} {13}},\ \bibinfo {pages} {2475}
  (\bibinfo {year} {2003})}\BibitemShut {NoStop}%
\bibitem [{\citenamefont {Melbinger}\ \emph {et~al.}(2012)\citenamefont
  {Melbinger}, \citenamefont {Reese},\ and\ \citenamefont
  {Frey}}]{melbinger2012}%
  \BibitemOpen
  \bibfield  {author} {\bibinfo {author} {\bibfnamefont {A.}~\bibnamefont
  {Melbinger}}, \bibinfo {author} {\bibfnamefont {L.}~\bibnamefont {Reese}}, \
  and\ \bibinfo {author} {\bibfnamefont {E.}~\bibnamefont {Frey}},\ }\href@noop
  {} {\bibfield  {journal} {\bibinfo  {journal} {Phys Rev Lett}\ }\textbf
  {\bibinfo {volume} {108}},\ \bibinfo {pages} {258104} (\bibinfo {year}
  {2012})}\BibitemShut {NoStop}%
\bibitem [{\citenamefont {Szavits-Nossan}\ \emph {et~al.}(2014)\citenamefont
  {Szavits-Nossan}, \citenamefont {Eden}, \citenamefont {Morris}, \citenamefont
  {MacPhee}, \citenamefont {Evans},\ and\ \citenamefont {Allen}}]{szavits2014}%
  \BibitemOpen
  \bibfield  {author} {\bibinfo {author} {\bibfnamefont {J.}~\bibnamefont
  {Szavits-Nossan}}, \bibinfo {author} {\bibfnamefont {K.}~\bibnamefont
  {Eden}}, \bibinfo {author} {\bibfnamefont {R.~J.}\ \bibnamefont {Morris}},
  \bibinfo {author} {\bibfnamefont {C.~E.}\ \bibnamefont {MacPhee}}, \bibinfo
  {author} {\bibfnamefont {M.~R.}\ \bibnamefont {Evans}}, \ and\ \bibinfo
  {author} {\bibfnamefont {R.~J.}\ \bibnamefont {Allen}},\ }\href@noop {}
  {\bibfield  {journal} {\bibinfo  {journal} {Phys Rev Lett}\ }\textbf
  {\bibinfo {volume} {113}},\ \bibinfo {pages} {098101} (\bibinfo {year}
  {2014})}\BibitemShut {NoStop}%
\bibitem [{\citenamefont {Rozhnova}\ and\ \citenamefont
  {Nunes}(2009)}]{rozhnova2009}%
  \BibitemOpen
  \bibfield  {author} {\bibinfo {author} {\bibfnamefont {G.}~\bibnamefont
  {Rozhnova}}\ and\ \bibinfo {author} {\bibfnamefont {A.}~\bibnamefont
  {Nunes}},\ }\href@noop {} {\bibfield  {journal} {\bibinfo  {journal} {Phys
  Rev E}\ }\textbf {\bibinfo {volume} {79}},\ \bibinfo {pages} {041922}
  (\bibinfo {year} {2009})}\BibitemShut {NoStop}%
\bibitem [{\citenamefont {Aoki}(2001)}]{aoki2001}%
  \BibitemOpen
  \bibfield  {author} {\bibinfo {author} {\bibfnamefont {M.}~\bibnamefont
  {Aoki}},\ }\href@noop {} {\emph {\bibinfo {title} {Modeling aggregate
  behavior and fluctuations in economics}}}\ (\bibinfo  {publisher} {Cambridge
  University Press, Cambridge},\ \bibinfo {year} {2001})\BibitemShut {NoStop}%
\bibitem [{\citenamefont {Heskes}(1994)}]{heskes1994}%
  \BibitemOpen
  \bibfield  {author} {\bibinfo {author} {\bibfnamefont {T.}~\bibnamefont
  {Heskes}},\ }\href@noop {} {\bibfield  {journal} {\bibinfo  {journal} {J Phys
  A: Math Gen}\ }\textbf {\bibinfo {volume} {27}},\ \bibinfo {pages} {5145}
  (\bibinfo {year} {1994})}\BibitemShut {NoStop}%
\bibitem [{\citenamefont {Grima}(2009)}]{grima2009}%
  \BibitemOpen
  \bibfield  {author} {\bibinfo {author} {\bibfnamefont {R.}~\bibnamefont
  {Grima}},\ }\href@noop {} {\bibfield  {journal} {\bibinfo  {journal} {Phys
  Rev Lett}\ }\textbf {\bibinfo {volume} {102}},\ \bibinfo {pages} {218103}
  (\bibinfo {year} {2009})}\BibitemShut {NoStop}%
\bibitem [{\citenamefont {Thomas}\ \emph {et~al.}(2010)\citenamefont {Thomas},
  \citenamefont {Straube},\ and\ \citenamefont {Grima}}]{thomas2010}%
  \BibitemOpen
  \bibfield  {author} {\bibinfo {author} {\bibfnamefont {P.}~\bibnamefont
  {Thomas}}, \bibinfo {author} {\bibfnamefont {A.~V.}\ \bibnamefont {Straube}},
  \ and\ \bibinfo {author} {\bibfnamefont {R.}~\bibnamefont {Grima}},\
  }\href@noop {} {\bibfield  {journal} {\bibinfo  {journal} {J Chem Phys}\
  }\textbf {\bibinfo {volume} {133}},\ \bibinfo {pages} {195101} (\bibinfo
  {year} {2010})}\BibitemShut {NoStop}%
\bibitem [{\citenamefont {Grima}\ \emph {et~al.}(2011)\citenamefont {Grima},
  \citenamefont {Thomas},\ and\ \citenamefont {Straube}}]{grima2011}%
  \BibitemOpen
  \bibfield  {author} {\bibinfo {author} {\bibfnamefont {R.}~\bibnamefont
  {Grima}}, \bibinfo {author} {\bibfnamefont {P.}~\bibnamefont {Thomas}}, \
  and\ \bibinfo {author} {\bibfnamefont {A.~V.}\ \bibnamefont {Straube}},\
  }\href@noop {} {\bibfield  {journal} {\bibinfo  {journal} {J Chem Phys}\
  }\textbf {\bibinfo {volume} {135}},\ \bibinfo {pages} {084103} (\bibinfo
  {year} {2011})}\BibitemShut {NoStop}%
\bibitem [{\citenamefont {Cianci}\ \emph {et~al.}(2012)\citenamefont {Cianci},
  \citenamefont {Di~Patti}, \citenamefont {Fanelli},\ and\ \citenamefont
  {Barletti}}]{cianci2012}%
  \BibitemOpen
  \bibfield  {author} {\bibinfo {author} {\bibfnamefont {C.}~\bibnamefont
  {Cianci}}, \bibinfo {author} {\bibfnamefont {F.}~\bibnamefont {Di~Patti}},
  \bibinfo {author} {\bibfnamefont {D.}~\bibnamefont {Fanelli}}, \ and\
  \bibinfo {author} {\bibfnamefont {L.}~\bibnamefont {Barletti}},\ }\href@noop
  {} {\bibfield  {journal} {\bibinfo  {journal} {Eur Phys J Special Topics}\
  }\textbf {\bibinfo {volume} {212}},\ \bibinfo {pages} {5} (\bibinfo {year}
  {2012})}\BibitemShut {NoStop}%
\bibitem [{\citenamefont {Thomas}\ \emph {et~al.}(2014)\citenamefont {Thomas},
  \citenamefont {Fleck}, \citenamefont {Grima},\ and\ \citenamefont
  {Popovic}}]{thomas2014}%
  \BibitemOpen
  \bibfield  {author} {\bibinfo {author} {\bibfnamefont {P.}~\bibnamefont
  {Thomas}}, \bibinfo {author} {\bibfnamefont {C.}~\bibnamefont {Fleck}},
  \bibinfo {author} {\bibfnamefont {R.}~\bibnamefont {Grima}}, \ and\ \bibinfo
  {author} {\bibfnamefont {N.}~\bibnamefont {Popovic}},\ }\href@noop {}
  {\bibfield  {journal} {\bibinfo  {journal} {J Phys A: Math Theor}\ }\textbf
  {\bibinfo {volume} {47}},\ \bibinfo {pages} {455007} (\bibinfo {year}
  {2014})}\BibitemShut {NoStop}%
\bibitem [{\citenamefont {Engblom}(2006)}]{engblom2006}%
  \BibitemOpen
  \bibfield  {author} {\bibinfo {author} {\bibfnamefont {S.}~\bibnamefont
  {Engblom}},\ }\href@noop {} {\bibfield  {journal} {\bibinfo  {journal} {Appl
  Math Comput}\ }\textbf {\bibinfo {volume} {180}},\ \bibinfo {pages} {498}
  (\bibinfo {year} {2006})}\BibitemShut {NoStop}%
\bibitem [{\citenamefont {Grima}(2012)}]{grima2012}%
  \BibitemOpen
  \bibfield  {author} {\bibinfo {author} {\bibfnamefont {R.}~\bibnamefont
  {Grima}},\ }\href@noop {} {\bibfield  {journal} {\bibinfo  {journal} {J Chem
  Phys}\ }\textbf {\bibinfo {volume} {136}},\ \bibinfo {pages} {154105}
  (\bibinfo {year} {2012})}\BibitemShut {NoStop}%
\bibitem [{\citenamefont {Ale}\ \emph {et~al.}(2013)\citenamefont {Ale},
  \citenamefont {Kirk},\ and\ \citenamefont {Stumpf}}]{ale2013}%
  \BibitemOpen
  \bibfield  {author} {\bibinfo {author} {\bibfnamefont {A.}~\bibnamefont
  {Ale}}, \bibinfo {author} {\bibfnamefont {P.}~\bibnamefont {Kirk}}, \ and\
  \bibinfo {author} {\bibfnamefont {M.~P.}\ \bibnamefont {Stumpf}},\
  }\href@noop {} {\bibfield  {journal} {\bibinfo  {journal} {J Chem Phys}\
  }\textbf {\bibinfo {volume} {138}},\ \bibinfo {pages} {174101} (\bibinfo
  {year} {2013})}\BibitemShut {NoStop}%
\bibitem [{\citenamefont {Sotiropoulos}\ and\ \citenamefont
  {Kaznessis}(2011)}]{sotiropoulos2011}%
  \BibitemOpen
  \bibfield  {author} {\bibinfo {author} {\bibfnamefont {V.}~\bibnamefont
  {Sotiropoulos}}\ and\ \bibinfo {author} {\bibfnamefont {Y.~N.}\ \bibnamefont
  {Kaznessis}},\ }\href@noop {} {\bibfield  {journal} {\bibinfo  {journal}
  {Chem Eng Sci}\ }\textbf {\bibinfo {volume} {66}},\ \bibinfo {pages} {268}
  (\bibinfo {year} {2011})}\BibitemShut {NoStop}%
\bibitem [{\citenamefont {Smadbeck}\ and\ \citenamefont
  {Kaznessis}(2013)}]{smadbeck2013}%
  \BibitemOpen
  \bibfield  {author} {\bibinfo {author} {\bibfnamefont {P.}~\bibnamefont
  {Smadbeck}}\ and\ \bibinfo {author} {\bibfnamefont {Y.~N.}\ \bibnamefont
  {Kaznessis}},\ }\href@noop {} {\bibfield  {journal} {\bibinfo  {journal}
  {Proc Natl Acad Sci}\ }\textbf {\bibinfo {volume} {110}},\ \bibinfo {pages}
  {14261} (\bibinfo {year} {2013})}\BibitemShut {NoStop}%
\bibitem [{\citenamefont {Andreychenko}\ \emph {et~al.}(2015)\citenamefont
  {Andreychenko}, \citenamefont {Mikeev},\ and\ \citenamefont
  {Wolf}}]{alexander2014}%
  \BibitemOpen
  \bibfield  {author} {\bibinfo {author} {\bibfnamefont {A.}~\bibnamefont
  {Andreychenko}}, \bibinfo {author} {\bibfnamefont {L.}~\bibnamefont
  {Mikeev}}, \ and\ \bibinfo {author} {\bibfnamefont {V.}~\bibnamefont
  {Wolf}},\ }\href@noop {} {\bibfield  {journal} {\bibinfo  {journal} {ACM
  Trans Model Comput Simul}\ }\textbf {\bibinfo {volume} {25}},\ \bibinfo
  {pages} {12} (\bibinfo {year} {2015})}\BibitemShut {NoStop}%
\bibitem [{\citenamefont {Cianci}\ \emph {et~al.}(2011)\citenamefont {Cianci},
  \citenamefont {Di~Patti},\ and\ \citenamefont {Fanelli}}]{cianci2011}%
  \BibitemOpen
  \bibfield  {author} {\bibinfo {author} {\bibfnamefont {C.}~\bibnamefont
  {Cianci}}, \bibinfo {author} {\bibfnamefont {F.}~\bibnamefont {Di~Patti}}, \
  and\ \bibinfo {author} {\bibfnamefont {D.}~\bibnamefont {Fanelli}},\
  }\href@noop {} {\bibfield  {journal} {\bibinfo  {journal} {Europhys Lett}\
  }\textbf {\bibinfo {volume} {96}},\ \bibinfo {pages} {50011} (\bibinfo {year}
  {2011})}\BibitemShut {NoStop}%
\bibitem [{\citenamefont {Cover}\ and\ \citenamefont
  {Thomas}(2012)}]{CoverThomas}%
  \BibitemOpen
  \bibfield  {author} {\bibinfo {author} {\bibfnamefont {T.~M.}\ \bibnamefont
  {Cover}}\ and\ \bibinfo {author} {\bibfnamefont {J.~A.}\ \bibnamefont
  {Thomas}},\ }\href@noop {} {\emph {\bibinfo {title} {Elements of information
  theory}}},\ \bibinfo {edition} {2nd}\ ed.\ (\bibinfo  {publisher} {John Wiley
  \& Sons, Hoboken, New Yersey},\ \bibinfo {year} {2012})\BibitemShut {NoStop}%
\bibitem [{\citenamefont {Thomas}\ \emph {et~al.}(2011)\citenamefont {Thomas},
  \citenamefont {Straube},\ and\ \citenamefont {Grima}}]{thomas2011}%
  \BibitemOpen
  \bibfield  {author} {\bibinfo {author} {\bibfnamefont {P.}~\bibnamefont
  {Thomas}}, \bibinfo {author} {\bibfnamefont {A.~V.}\ \bibnamefont {Straube}},
  \ and\ \bibinfo {author} {\bibfnamefont {R.}~\bibnamefont {Grima}},\
  }\href@noop {} {\bibfield  {journal} {\bibinfo  {journal} {J Chem Phys}\
  }\textbf {\bibinfo {volume} {135}},\ \bibinfo {pages} {181103} (\bibinfo
  {year} {2011})}\BibitemShut {NoStop}%
\bibitem [{\citenamefont {Sanft}\ \emph {et~al.}(2011)\citenamefont {Sanft},
  \citenamefont {Gillespie},\ and\ \citenamefont {Petzold}}]{sanft2011}%
  \BibitemOpen
  \bibfield  {author} {\bibinfo {author} {\bibfnamefont {K.~R.}\ \bibnamefont
  {Sanft}}, \bibinfo {author} {\bibfnamefont {D.~T.}\ \bibnamefont
  {Gillespie}}, \ and\ \bibinfo {author} {\bibfnamefont {L.~R.}\ \bibnamefont
  {Petzold}},\ }\href@noop {} {\bibfield  {journal} {\bibinfo  {journal} {IET
  Syst Biol}\ }\textbf {\bibinfo {volume} {5}},\ \bibinfo {pages} {58}
  (\bibinfo {year} {2011})}\BibitemShut {NoStop}%
\bibitem [{\citenamefont {Paulsson}\ \emph {et~al.}(2000)\citenamefont
  {Paulsson}, \citenamefont {Berg},\ and\ \citenamefont
  {Ehrenberg}}]{paulsson2000}%
  \BibitemOpen
  \bibfield  {author} {\bibinfo {author} {\bibfnamefont {J.}~\bibnamefont
  {Paulsson}}, \bibinfo {author} {\bibfnamefont {O.~G.}\ \bibnamefont {Berg}},
  \ and\ \bibinfo {author} {\bibfnamefont {M.}~\bibnamefont {Ehrenberg}},\
  }\href@noop {} {\bibfield  {journal} {\bibinfo  {journal} {Proc Natl Acad
  Sci}\ }\textbf {\bibinfo {volume} {97}},\ \bibinfo {pages} {7148} (\bibinfo
  {year} {2000})}\BibitemShut {NoStop}%
\bibitem [{\citenamefont {Comtet}(1974)}]{comtet1974}%
  \BibitemOpen
  \bibfield  {author} {\bibinfo {author} {\bibfnamefont {L.}~\bibnamefont
  {Comtet}},\ }\href@noop {} {\emph {\bibinfo {title} {Advanced
  Combinatorics}}}\ (\bibinfo  {publisher} {Springer, Netherlands},\ \bibinfo
  {year} {1974})\BibitemShut {NoStop}%
\bibitem [{Note1()}]{Note1}%
  \BibitemOpen
  \bibinfo {note} {The partial~Bell polynomials are defined as
  $B_{n,k}(\protect \{x_\chi\protect \})= {\DOTSB \sum@ \slimits@ }' {n!
  \protect \primfrac {over}j_1!.. j_{n-k+1}!} \left ({x_1\protect \primfrac
  {over}1!}\right )^{j_1}..\left ({x_{n-k+1} \protect \primfrac
  {over}(n-k+1)!}\right )^{j_{n-k+1}}, $ where the summation $\DOTSB \sum@
  \slimits@ '$ is such that $j_1 + \protect \ldots + j_{n-k+1}=k$ and $j_1 + 2
  j_2 + \protect \ldots + ({n-k+1})j_{n-k+1}=n$ and $\protect \{x_\chi\protect
  \}$ denotes the sequence $(x_1,x_2,x_3,\ldots)$. These are available in
  Mathematica via the function \protect \text {BellY}}\BibitemShut {NoStop}%
\bibitem [{\citenamefont {Thomas}\ \emph {et~al.}(2012)\citenamefont {Thomas},
  \citenamefont {Matuschek},\ and\ \citenamefont {Grima}}]{thomas2012}%
  \BibitemOpen
  \bibfield  {author} {\bibinfo {author} {\bibfnamefont {P.}~\bibnamefont
  {Thomas}}, \bibinfo {author} {\bibfnamefont {H.}~\bibnamefont {Matuschek}}, \
  and\ \bibinfo {author} {\bibfnamefont {R.}~\bibnamefont {Grima}},\ }
  Computation of biochemical pathway fluctuations beyond the linear noise approximation using iNA.
  in\
  \href@noop {} {\emph {\bibinfo {booktitle} {Bioinformatics and Biomedicine
  (BIBM), 2012 IEEE International Conference on}}}\ (\bibinfo {organization}
  {IEEE},\ \bibinfo {year} {2012})\ pp.\ \bibinfo {pages} {1--5}\BibitemShut
  {NoStop}%
\bibitem [{\citenamefont {Shahrezaei}\ and\ \citenamefont
  {Swain}(2008)}]{shahrezaei2008}%
  \BibitemOpen
  \bibfield  {author} {\bibinfo {author} {\bibfnamefont {V.}~\bibnamefont
  {Shahrezaei}}\ and\ \bibinfo {author} {\bibfnamefont {P.~S.}\ \bibnamefont
  {Swain}},\ }\href@noop {} {\bibfield  {journal} {\bibinfo  {journal} {Proc
  Natl Acad Sci}\ }\textbf {\bibinfo {volume} {105}},\ \bibinfo {pages} {17256}
  (\bibinfo {year} {2008})}\BibitemShut {NoStop}%
\bibitem [{Note2()}]{Note2}%
  \BibitemOpen
  \bibinfo {note} {The polylogarithm is defined by $\protect \operatorname
  {Li}_s(x) = \DOTSB \sum@ \slimits@ _{k=1}^\infty {x^k \protect \primfrac
  {over}k^s}$ and implemeted in Mathematica by PolyLog[$x,s$].}\BibitemShut
  {Stop}%
\bibitem [{\citenamefont {Gillespie}(1977)}]{gillespie1977}%
  \BibitemOpen
  \bibfield  {author} {\bibinfo {author} {\bibfnamefont {D.~T.}\ \bibnamefont
  {Gillespie}},\ }\href@noop {} {\bibfield  {journal} {\bibinfo  {journal} {J
  Phys Chem}\ }\textbf {\bibinfo {volume} {81}},\ \bibinfo {pages} {2340}
  (\bibinfo {year} {1977})}\BibitemShut {NoStop}%
\bibitem [{Note3()}]{Note3}%
  \BibitemOpen
  \bibinfo {note} {Exact realizations of the process described by Eq.
  (\ref{appl:burstapprox}) are obtained using the Gillespie's algorithm with
  the increments of the synthesis reaction being replaced by independently and
  identically geometrically distributed integers of mean $b$.}\BibitemShut
  {Stop}%
\bibitem [{\citenamefont {Verhulst}(2006)}]{verhulst2006}%
  \BibitemOpen
  \bibfield  {author} {\bibinfo {author} {\bibfnamefont {F.}~\bibnamefont
  {Verhulst}},\ }\href@noop {} {\emph {\bibinfo {title} {Methods and
  applications of singular perturbations}}}\ (\bibinfo  {publisher} {Springer,
  New York},\ \bibinfo {year} {2006})\BibitemShut {NoStop}%
\bibitem [{\citenamefont {Munsky}\ and\ \citenamefont
  {Khammash}(2006)}]{munsky2006}%
  \BibitemOpen
  \bibfield  {author} {\bibinfo {author} {\bibfnamefont {B.}~\bibnamefont
  {Munsky}}\ and\ \bibinfo {author} {\bibfnamefont {M.}~\bibnamefont
  {Khammash}},\ }\href@noop {} {\bibfield  {journal} {\bibinfo  {journal} {J
  Chem Phys}\ }\textbf {\bibinfo {volume} {124}},\ \bibinfo {pages} {044104}
  (\bibinfo {year} {2006})}\BibitemShut {NoStop}%
\bibitem [{\citenamefont {Lebedev}(1972)}]{Lebedev}%
  \BibitemOpen
  \bibfield  {author} {\bibinfo {author} {\bibfnamefont {N.~N.}\ \bibnamefont
  {Lebedev}},\ }\href@noop {} {\emph {\bibinfo {title} {Special functions and
  their applications}}}\ (\bibinfo  {publisher} {Dover Publications, New
  York},\ \bibinfo {year} {1972})\BibitemShut {NoStop}%
\end{thebibliography}
\end{document}